\documentclass[aps,prb,twocolumn,epsf,floatfix]{revtex4}
\usepackage{graphicx}
\usepackage{latexsym}
\usepackage{amsmath}
\begin{document}
\title{Superfluid-Insulator Transitions on the Triangular Lattice}
\author{A.A. Burkov and Leon Balents} \affiliation{\small Department
  of Physics, University of California, Santa Barbara, CA 93106}
\date{\today}

\begin{abstract}
  We report on a phenomenological study of superfluid to Mott
  insulator transitions of bosons on the triangular lattice, focusing
  primarily on the interplay between Mott localization and geometrical
  charge frustration at 1/2-filling.  A general dual vortex field
  theory is developed for arbitrary rational filling factors $f$,
  based on the appropriate projective symmetry group.  At the simple
  non-frustrated density $f=1/3$, we uncover an example of a
  deconfined quantum critical point very similar to that found on the
  half-filled square lattice.  Turning to $f=1/2$, the behavior is
  quite different.  Here, we find that the low-energy action
  describing the Mott transition has an emergent nonabelian $SU(2)
  \times U(1)$ symmetry, not present at the microscopic level.  This
  large nonabelian symmetry is directly related to the
  frustration-induced quasi-degeneracy between many charge-ordered states not
  related by microscopic symmetries.  Through this ``pseudospin''
  $SU(2)$ symmetry, the charged excitations in the insulator close to
  the Mott transition develop a skyrmion-like character.  This leads
  to an understanding of the recently discovered supersolid phase of
  the triangular lattice XXZ model\cite{Melko05,Damle05,Troyer05} as a
  ``partially melted'' Mott insulator.  The latter picture naturally
  explains a number of puzzling numerical observations of the
  properties of this supersolid.  Moreover, we predict that the nearby
  quantum phase transition from this supersolid to the Mott insulator
  is in the recently-discovered non-compact CP$^1$
  critical universality class.\cite{Lesik}\  A description of a broad range of
  other Mott and supersolid states, and a diverse set of quantum
  critical points between them, is also provided.
\end{abstract}

\maketitle
\section{Introduction}
\label{sec:intro}

Theoretical interest in quantum phase transitions from superfluid to
Mott insulating states of bosons has recently been revived by their
experimental realization in cold atoms trapped in an optical
lattice.\cite{Greiner} Extensions of these experiments are expected to
soon provide a great variety of real life toy models, where
theoretical scenarios for such phase transitions can be
tested.\cite{Demler03,Zoller04,Zoller05} 
For the condensed matter community, such
transitions are interesting in their own right, but also provide a
simpler context in which some aspects of Mott conducting-insulating
transitions of {\sl electrons} can be explored.  A better
understanding of such Mott criticality generally may help to explain
mysteries in various strongly correlated materials, from heavy fermion
metals \cite{livrefs} to cuprate superconductors,\cite{cuprefs} in which
Mott criticality may plausibly be argued to play a key role.  

An exciting theoretical development in the field has been the
discovery that some quantum phase transitions require a fundamentally
new description, not based on the now-standard Landau's concept of an
order parameter.  Instead, such quantum critical points (QCPs) are described in
terms of {\em emergent} degrees of freedom, not present in either
phase and appearing due to certain special dynamically generated
low-energy symmetries at the critical point.\cite{dcprefs} An
interesting consequence of the emergent low-energy symmetry of the
critical point is the near-degeneracy of ``competing ordered'' states
unrelated by any {\sl microscopic} symmetry (but unified with one
another by the emergent one) in the neighborhood of the quantum phase
transition.  There is, at present, unfortunately, no general way to
{\sl a priori} identify the appropriate emergent degrees of freedom,
should they exist, for any putative quantum critical point.

In the particular context of two-dimensional bosonic
superfluid-insulator transitions, a general non-Landau-Ginzburg-Wilson
(non-LGW) framework {\sl has} recently been proposed in
Ref.~\onlinecite{Balents04} (and see Ref.~\onlinecite{ykis} for a
pedagogical review), and carried out explicitly on the square lattice.
In particular, the Mott transition can be described in terms of the
{\sl vortex} excitations of the superfluid.  In a two dimensional
superfluid, vortices are point-like ``particles'' whose
creation/annihilation operators can be used to construct a quantum
field theory.  The vortices being non-local topological objects, these
vortex fields are, however, not themselves order parameters in
the LGW sense.  The non-locality is manifested by the presence of a
(non-compact) $U(1)$ gauge field, to which the vortex fields are
coupled in the ``dual'' vortex field theory.  This formulation is
general because it is based on the excitations of the {\sl
  superfluid}, which is a stable and apparently featureless (i.e.
without broken symmetry apart from off-diagonal long range order)
state at any boson density.  Nevertheless, it was shown in
Ref.~\onlinecite{Balents04} that the vortices exhibit a subtle {\em
  quantum order} which {\sl is} sensitive to the boson density $f$
(per unit cell of the lattice).  In particular, at non-integral 
$f$, the vortices form non-trivial {\sl multiplets}
transforming under a {\sl projective symmetry group} (PSG -
technically, a projective representation of the lattice space group).
The Lagrange density for the vortex field theory therefore takes the
general form
\begin{eqnarray}
  \label{eq:vft}
  {\cal L} &=&\sum_{\ell = 0}^{N}
\left[|(\partial_{\mu} - i A_{\mu}) \varphi_{\ell}|^2 + 
s |\varphi_{\ell}|^2 \right]+ {\cal L}_{\rm int}[\{\varphi_\ell\}]
\nonumber \\  
&&+ \frac{1}{2 e^2} 
(\epsilon_{\mu \nu \lambda}\partial_{\nu} A_{\lambda})^2,
\end{eqnarray}
where $\varphi_\ell$ with $\ell=1\ldots N$ are vortex fields for the
$N$ members of the multiplet ($N$ depends upon $f$ -- see
Sec.~\ref{sec:dual}), and $A_\mu$ is the dual $U(1)$ gauge field.
Quartic and higher order terms are contained in ${\cal L}_{\rm int}$,
the structure of which is dictated by the PSG.  The Mott transition is
captured by Eq.(\ref{eq:vft}) in a simple way.  The ground state of
$s > 0$ corresponds to the vacuum of vortices -- flux is expelled from
the system, so this is a superfluid.  In the $s < 0$ phase, at least
one of the vortex flavors will condense, and the gauge fields will
acquire a gap by the Higgs mechanism.  This is indicative of a charge
gap, and describes an {\em incompressible} Mott insulating phase.

Transformations within the $\varphi_\ell$ multiplet comprise the
emergent symmetry operations of the ``deconfined'' quantum critical
point.  This structure has an important physical consequence as well:
it naturally and unavoidably leads to (particular) broken spatial
symmetries in the Mott state.  A mean-field analysis of the vortex
field theory predicts a direct superfluid to Mott transition, as well
as the nature of the charge ordering in the Mott phase.

In this paper, we extend this vortex field theory approach to describe
superfluid to Mott insulator transitions of bosons on the triangular
lattice ({\sl hexagonal} lattice in proper crystallographic
nomenclature) at fractional boson fillings $f=p/q$, with $p,q$
relatively prime integers.  We focus particularly on the most
interesting examples, $f=1/2$ and $f=1/3$.  The $1/2$-filling case
introduces a new ingredient not present on the square lattice:
geometrical frustration.  Here we refer to ``charge frustration'' of
the ordering of localized boson configurations in the presence of
short-range repulsive interactions.  A consequence -- or perhaps
definition -- of such geometrical frustration is the exact or near
degeneracy of many distinct ordered states.  The similarity of this
property with the emergent near-degeneracy of competing orders near a
deconfined quantum critical point suggests a possible link between the
two phenomena.  This connection indeed appears to be borne out  by the analysis in this paper.

In the simplest classical models of frustration, the degeneracy
amongst low-energy states is not only large but macroscopic (i.e. with
an entropy proportional to the sample volume).  This classical
degeneracy, when lifted by quantum fluctuations, may produce unusual
ground states.  A number of very recent
papers\cite{Melko05,Damle05,Troyer05}\ have investigated the system of
hardcore bosons with nearest-neighbor interactions (which can be
mapped to a spin-$1/2$ XXZ model) on the triangular
lattice near half-filling.  This realizes such approximately classical
frustration in the limit of very strong near-neighbor repulsion.
These works demonstrated that in this system the lifting of the
macroscopic classical degeneracy results in an unusual {\em
  supersolid} ground state, which we denote SS3 because of its
3-sublattice structure.  This ``order by disorder'' mechanism is
very different from the ``conventional'' (theoretically!)  picture of
supersolidity, via a condensation of
vacancies and/or interstitials in an ordered solid.\cite{supersolid}

Since the above studies clarified that such Mott states do not occur
in the simplest nearest-neighbor interaction model, it is apparent
that microscopically, longer-range interactions, possibly including
ring-exchange,\cite{Sandvik02} are necessary to observe these
transitions in a microscopic model.  Finding simple interactions that
produce nontrivial insulating ground states on the triangular
lattice is an important and difficult problem, that will likely
require sophisticated numerical methods.  We will not address this
issue here (but see the Discussion, Sec.~\ref{fin}).

Our phenomenological vortex field theory, on the contrary, describes
universal aspects (independent of microscopic realization) of Mott
insulating, superfluid, and other states, and the transitions between
them.  As for the square lattice case, a mean-field analysis predicts
a direct superfluid-Mott transition, with a diverse set of Mott
insulating phases.  Also like the square lattice, the vortex field
theory has an enhanced emergent symmetry at the critical point, a
hallmark of deconfined criticality.  A significant difference from
these prior examples of deconfined criticality, however, is that, in
the frustrated case, $f=1/2$, the emergent symmetry is {\sl
  nonabelian}, containing an $SU(2)$ ``pseudospin'' subgroup.  The
much larger (than in non-frustrated cases) emergent symmetry can be
understood physically as symptomatic of the larger near-degeneracies
present in this case due to frustration.

Remarkably, going beyond the mean-field analysis of the vortex field
theory, our approach connects very nicely to the supersolid phase of
the XXZ model.  Indeed, the supersolid order parameter -- describing
the growth of the supersolid state out of the featureless superfluid
-- appears in a particularly simple form in the vortex variables.  
Moreover, our approach reveals an alternative view of the supersolid,
as a partially-melted ``parent'' Mott insulating state, with ``quantum
disordered'' pseudospin.  This loss of pseudospin order
{\sl simultaneously} with the onset of superfluidity is possible
because, as we show, the pseudospin {\sl skyrmion} excitation of the
Mott insulator carries physical boson charge.  The
supersolid may thereby also be viewed as a condensate of these
skyrmions.   Furthermore, this picture leads directly to the
prediction that the Mott insulator to SS3 transition is described by
the recently discovered Non-Compact CP$^1$ (NCCP$^1$) quantum critical
universality class.\cite{Lesik}\ This transition describes the quantum
disordering of a pseudospin vector ${\bf S}$ (the order parameter for
the additional solid order of the Mott state) in $2+1$ dimensions,
when ``hedgehog'' instantons are absent in space-time.  These hedgehog
events correspond precisely to processes which change the skyrmion
number, and are therefore prohibited by charge conservation in this
case.   

The paper is organized as follows.  In Sec.\ref{sec:dual} we
develop the dual vortex theory for the triangular lattice at a
general rational filling. We derive the PSG transformations of the
degenerate low energy vortex modes and discuss some of their general
properties.  In Sec.~\ref{mft} we apply this theory to the two cases
$f=1/3$ and $f=1/2$, and discuss the resulting Mott states that are
obtained by a mean field analysis of the vortex field theory.  In
Sec.~\ref{defects} we present a ``hard-spin'' formulation of the
vortex action, which enables a study of its phases and transitions
beyond mean field theory.  We discuss the new phases which arise,
notably supersolids, the elementary excitations of the different
states, and the transitions amongst them.  This includes notably the
SS3 phases and the NCCP$^1$ transitions between them and their parent
Mott insulator. We conclude with a
discussion in Sec.~\ref{fin} of a more microscopic physical picture
for the most interesting Mott and supersolid states at half-filling,
the connection to the recent studies of the XXZ model, and the
prospects for observing more of the physics in this paper in related
models.  A number of appendices provide useful details of various technical
results.

\section{Continuum dual vortex theory}
\label{sec:dual}

As discussed in the introduction, our aim is to derive a field
theoretic description of the vicinity of a superfluid to Mott
insulator transition, in terms of the vortex excitations of the
superfluid state.  As is well-known, lattice models of bosons can
indeed be reformulated in vortex variables on the dual lattice, a
technique called duality.  As detailed for instance in
Ref.~\onlinecite{ykis} (and references therein), this can in principle be
carried out exactly for any lattice boson Hamiltonian, e.g.
Bose-Hubbard models, XXZ models, etc.  Unfortunately, this exact
mapping does not lead to a particularly tractable limit of the vortex
theory, and it is therefore difficult to extract quantitative
predictions directly from the lattice vortex theory.  

Fortunately, we can avoid this difficulty in addressing our stated
goal of understanding universal phenomena in the vicinity of the
superfluid-Mott quantum critical point.  For this purpose, we need not
specify any particular microscopic boson model.  Instead, we will
illustrate the calculations through the use of a very simple lattice
vortex theory, which is chosen to have the same spatial symmetries
as the physical triangular lattice, and to exhibit a superfluid to
Mott insulation transition.  The universal properties of interest will
coincide with those of more physical microscopic boson models.  

In the Euclidean vortex coherent-state path integral
formulation in which we work, the lattice vortex action is
\begin{eqnarray}
\label{eq:3}
S&=&-t_v \sum_{a\mu}[\psi^*_{a} e^{-i A_{a\mu}} \psi_{a+\mu} + c.c.] 
\nonumber \\ 
&+&\sum_a [s |\psi_a|^2 + u |\psi_a|^4] \nonumber \\  
&+&\frac{1}{2 e^2} \sum
\left(\epsilon_{\mu \nu \lambda} \Delta_{\nu} A_{a \lambda} - 2 \pi f 
\delta_{\mu \tau}\right)^2.  
\end{eqnarray} 
Here $a$ labels sites of the dual 2+1-dimensional uniformly stacked
honeycomb lattice, with the vertical direction being the imaginary
time direction $\tau$, and $\mu$ is summed over nearest-neighbor links
(3 at $120$ degrees connecting spatial neighbors and a fourth along
the imaginary time direction).  The action is written in terms of the complex
vortex field $\psi_a$ (and its conjugate $\psi_a^*$) which annihilates
(creates) unit vorticity, as well as a dual gauge field $A_{a\mu}$.
The physical meaning of the gauge field $A_{a\mu}$ is that its curl
gives ($2\pi$ times) the bosonic 3-current density, and importantly,
the temporal component of the current density is ($2\pi$ times) the
charge density.  For a bosonic system with density $f$ bosons per
site, we must therefore enforce the condition that $2\pi f$ flux
on average passes through each hexagonal plaquette of the honeycomb
lattice.  We will assume the filling to
be rational $f = p/q$, where $p$ and $q$ are relatively prime
integers, and will mostly concentrate on the two cases $f=1/2, 1/3$.
Physically, $t_v$ represents a vortex hopping amplitude, $s$ and $u$
represent short-range vortex ``core'' energies and interactions, and $e^2$
represents the strength of the dual electromagnetic field fluctuations
(it is roughly proportional to the local superfluid density
away from the Mott quantum critical point).

\begin{figure}[t]
\includegraphics[width=6cm]{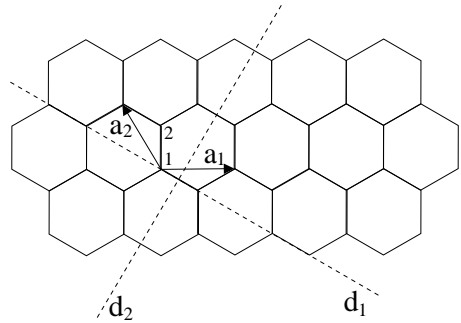}
\caption{Dual honeycomb lattice. Reflection axes $d_1$ and $d_2$ are 
shown by dashed lines.} 
\label{hexlat}
\end{figure}

We may think of the parameter $s$ in Eq.(\ref{eq:3}) as driving the
superfluid-insulator transition. Large positive $s$ corresponds to the
superfluid phase, in which vortices are gapped, and large negative $s$
corresponds to the set of possible insulating phases, in which
vortices are condensed.  It is convenient to think about the
transition in Eq.~(\ref{eq:3}) for small $e^2$, i.e. neglecting to a
first approximation dual gauge fluctuations -- they will however be
restored at a later stage of analysis.  In this limit we can treat the
dual gauge field in a mean field approximation and take:
\begin{equation}
\label{eq:4}
\epsilon_{\mu \nu \lambda} \Delta_{\nu} A_{a\lambda} = 2 \pi f 
\delta_{\mu \tau}.
\end{equation}
As $s$ is decreased from large positive values, we expect an
instability of the ``vortex vacuum'' when the energy of the lowest
vortex excitation approaches zero -- or equivalently, when the minimum
eigenvalue of the quadratic form for $\psi_a,\psi_a^*$ in
Eq.~(\ref{eq:3}) vanishes.  To study the universal critical properties
of the superfluid-insulator transition, it is sufficient to isolate
these low-energy particles, which comprise the vortex multiplets
discussed in the introduction.  The continuum limit of
Eq.~(\ref{eq:3}) then consists of a set of vortex fields, each
representing one of these minimum energy vortex particles. We will
also take the trivial continuum limit of Eq.(\ref{eq:3}) in the
temporal direction.

Because of the non-zero gauge flux through each spatial plaquette, the
minimum energy multiplets are non-trivial.  The form of the continuum
action in this case is determined by the projective representation of
the space group (PSG), \cite{Balents04} under which the vortex fields
$\psi$ transform.  To find it, we must work through the consequences
of some specific gauge choice.  As in Ref.\onlinecite{Balents04}, we
will choose the Landau gauge for $A_{a \mu}$.  Namely, let
\begin{equation}
\label{eq:5}
{\bf a}_1 = \hat x, \,\,\, {\bf a}_2 = -\frac{1}{2} \hat x + \frac{\sqrt{3}}
{2} \hat y,
\end{equation}
be the two basis vectors of the honeycomb lattice, as shown in 
Fig.\ref{hexlat} (note that the honeycomb lattice has two sites per
unit cell).  Coordinates will be specified, when explicit, in this
basis, ${\bf r} = a_1 {\bf a}_1 + a_2 {\bf a}_2$, with integer
$a_1,a_2$, and $a_1=a_2=0$ corresponding to a ``type 1'' site (see
Fig.~\ref{hexlat}) of the dual honeycomb lattice.  
Landau gauge is
\begin{equation}
\label{eq:6}
A_{ay} = 2 \pi f a_1,
\end{equation}
and $A_{a\mu} = 0$ for all other directions; that is, only the gauge
field on vertical links is chosen non-zero.  The full PSG is generated
by a set of unitary transformations of $\psi$, one for each generator of
the lattice space group.  For the triangular lattice, we choose the
space group generators as two elementary lattice translations $T_1,T_2$,
a $2\pi/3$ rotation with respect to site $1$ in Fig.~\ref{hexlat},
$R_{2\pi/3}$, and two reflections, $I_{d_1},I_{d_2}$.\footnote{It is
  straightforward to show that the $6$-fold rotation about a direct
  lattice site, $R_6$, is determined from these generators by the
  relation $R_6 \circ T_2 \circ R_{2\pi/3}\circ I_{d_1}\circ I_{d_2}=1$,
  and so is not independent.}  The PSG transformations of the vortex
fields, characterized by the unimodular complex number $\omega = e^{2
  \pi i f}$, are then:
\begin{eqnarray}
\label{eq:7}
T_1&:& \left\{ \begin{array}{lcl} \psi_1(a_1,a_2) & \rightarrow &
    \psi_1(a_1\!-\!1,a_2) \omega^{a_2} \\ 
 \psi_2(a_1,a_2) & \rightarrow & \psi_2(a_1\!-\!1,a_2) \omega^{a_1+1}
         \end{array} \right.  ,\nonumber \\
T_2&:&\left\{ \begin{array}{lcl} \psi_1(a_1,a_2) & \rightarrow & \psi_1(a_1,a_2\!-\!1) \\
\psi_2(a_1,a_2) & \rightarrow & \psi_2(a_1,a_2\!-\!1)\end{array}
\right. ,\nonumber \\ 
R_{2\pi/3}&:& \left\{ \begin{array}{lcl} \psi_1(a_1,a_2) & \rightarrow
    & \psi_1(a_2\!-\!a_1,-a_1) 
\omega^{\frac{a_1(2a_2-a_1-1)}{2}} \\
\psi_2(a_1,a_2) & \rightarrow&  \psi_2(a_2\!-\!a_1,-a_1\!-\!1) 
\omega^{\frac{a_1(2a_2-a_1+1)}{2}}\end{array}\right. , \nonumber \\
I_{d_1}&:& \left\{ \begin{array}{lcl} \psi_1(a_1,a_2) & \rightarrow &
    \psi_1^*(-a_2,-a_1) \omega^{a_1 a_2} \\
\psi_2(a_1,a_2) & \rightarrow & \psi_2^*(-a_2\!-\!1,-a_1\!-\!1) 
\omega^{a_1(a_2+1)}\end{array}\right. , \nonumber \\
I_{d_2}&:& \left\{ \begin{array}{lcl} \psi_1(a_1,a_2) & \rightarrow &  \psi_2^*(a_2,a_1\!-\!1) \omega^{a_1 a_2} \\
\psi_2(a_1,a_2)& \rightarrow & \psi_1^*(a_2\!+\!1,a_1) 
\omega^{a_1(a_2+1)}\end{array}\right. .
\end{eqnarray}
For the special case $f=1/2$, in which we are principally interested,
one usually considers in addition to these spatial symmetries, an extra
{\sl particle-hole symmetry}, which we denote $C$ for ``charge
conjugation''.  This interchanges singly-occupied and empty sites on the
direct lattice.  In the spin language appropriate for hard-core bosons,
this is just a $180^\circ$ rotation around the $x$ axis in spin space,
which has the effect of an Ising transformation $S_i^z \rightarrow
-S_i^z$ (and likewise for $S_i^y$).  The XXZ model, and indeed any hard
core boson model at $f=1/2$ with only pairwise interactions, possesses
such a particle-hole symmetry.  In the dual theory, this requires the
action to be invariant under
\begin{equation}
  \label{eq:ph}
 C \, : \,  \psi_i(a_1,a_2) \rightarrow \psi_i^*(a_1,a_2),
\end{equation}
and simultaneous sign change  of the fluctuating part of the dual gauge
field $A_\mu \rightarrow - A_\mu$.

The quadratic action of Eq.~(\ref{eq:3}) in Landau gauge has a
periodicity in real space of $q$ unit cells in the ${\bf a}_1$
direction, and one unit in the ${\bf a}_2$ direction (note that, of
course, the physics itself has the full periodicity of the honeycomb
(or underlying triangular direct) lattice).  The eigenstates of the
quadratic action can therefore be characterized by their quasimomenta
in the corresponding reduced Brillouin zone.  Specifically, we introduce 
the basis vectors of the reciprocal lattice,
\begin{equation}
\label{eq:8}
{\bf b}_1 = \hat x + \frac{1}{\sqrt{3}} \hat y, 
\,\,\, {\bf b}_2 = \frac{2}{\sqrt{3}} \hat y.
\end{equation}
Wavevectors will, when necessary, be specified by coordinates
$(k_1,k_2)$, with ${\bf k}=k_1 {\bf b}_1+k_2 {\bf b_2}$ (reciprocal
lattice vectors correspond to $k_1,k_2$ being integral multiples of $2\pi$).  

By applying the methods of Ref.\onlinecite{Balents04}, one may readily find
the minimum energy multiplet for arbitrary $p,q$, and their PSG
transformations.  Briefly, this is accomplished by Fourier
transforming Eqs.~(\ref{eq:7}) to obtain the PSG for general
$\psi({\bf k})$ in momentum space -- this is given in
Appendix~\ref{sec:psg-momentum-space} -- and using the non-commutative
algebra of translations and rotations implicit in Eqs.~(\ref{eq:7}) to
generate a full set of eigenfunctions.  
One finds two cases.  For $q$ odd, there are $q$ minima of the vortex
dispersion, i.e. $N=q$ in Eq.~(\ref{eq:vft}).  These occur at momenta
\begin{equation}
\label{eq:11}
{\bf k}_{\ell} = \left(0, 2 \pi f \ell\right),\,\,\, \ell= 0,\ldots,q-1,
\end{equation}
taking wavevectors ${\bf k}_\ell$ to lie in the reduced Brilloin zone
$-\pi/q \le k_1 < \pi/q$ and $-\pi \le k_2 < \pi$.  By contrast, for
$q$ even, $N=2q$, and it is convenient to divide minima into two sets
$\alpha=\pm 1 \equiv \pm$, parameterizing $\ell=(1+\alpha)q/2 +
\sigma$, $\sigma=0\ldots q-1$.  The vortex field
operator $\varphi_{\alpha\sigma}$ then acts on eigenstates with the
wavevector
\begin{equation}
\label{eq:10}
{\bf k}_{\alpha\sigma} = \left(\pi \alpha / 3 q,
\pi \alpha / 3 q + 2 \pi f \sigma\right)
\end{equation}
in the reduced Brillouin zone.  For even $q$, therefore,
Eq.~(\ref{eq:vft}) may be rewritten as 
\begin{eqnarray}
  \label{eq:vfteven}
{\cal L}_0^{even}&=&\sum_{\alpha=\pm} \sum_{\sigma = 0}^{q-1}
\left[|(\partial_{\mu} - i A_{\mu}) \varphi_{\alpha\sigma}|^2 + 
s |\varphi_{\alpha\sigma}|^2 \right] \nonumber \\ 
&+&\frac{1}{2 e^2} 
(\epsilon_{\mu \nu \lambda}\partial_{\nu} A_{\lambda})^2.
\end{eqnarray}
We note in passing that the set of
$\varphi_\ell$ corresponding to these wavevectors gives the smallest
number of minima possible for the vortex dispersion (i.e. they
comprise the smallest-dimensional representation of the PSG for the
given $f$), and they are what is realized without special fine-tuning
of the vortex kinetic energy terms.

The PSG transformations of the multiplet can be found using
Eq.(\ref{eq:9}) of Appendix~\ref{sec:psg-momentum-space}, by a
straightforward generalization of the procedure, detailed in
Ref.~\onlinecite{Balents04}.  For instance, under the translations,
one finds
\begin{eqnarray}
  \label{eq:PSGtodd}
  T_1 : && \varphi_{\ell} \rightarrow \varphi_{\ell-1}, \nonumber \\
  T_2 : &&\varphi_{\ell} \rightarrow \omega^{-\ell} \varphi_{\ell},   
\end{eqnarray}
for $q$ odd, and
\begin{eqnarray}
  \label{eq:PSGteven}
  T_1 : && \varphi_{\alpha\sigma} \rightarrow e^{-i \pi \alpha/3 q}
  \varphi_{\alpha,\sigma-1}, \nonumber \\
  T_2 : && \varphi_{\alpha\sigma} \rightarrow e^{-i \pi \alpha/3 q} 
  \omega^{-\sigma} \varphi_{\alpha\sigma},
\end{eqnarray}
for $q$ even.  Here and in the following, the index $\ell$ for odd $q$
and $\sigma$ for even $q$ will be regarded as cyclic modulo $q$.  The
remaing PSG generators are given in Eqs.~(\ref{eq:14},\ref{eq:12}) in
Appendix~\ref{sec:psg-momentum-space}.  

It is now straightforward to write down the most general continuum
vortex Lagrangian density, describing the superfluid-insulator
transition on the triangular lattice.  The most general terms at
quadratic order are simply given by Eqs.~(\ref{eq:vft},\ref{eq:vfteven}).
Let us now consider the quartic potential terms in the continuum theory. 
Following the general approach of Ref.~\onlinecite{Balents04}, we first 
write down the continuum Lagrangian, imposing only the restrictions from 
gauge symmetry and translational symmetry, i.e. invariance under
Eqs.~(\ref{eq:PSGtodd},\ref{eq:PSGteven}).  One finds
\begin{eqnarray}
\label{eq:16}
{\cal L}_1^{\rm odd} & = & \sum_{\ell mn}^{q-1} 
\gamma_{mn} \varphi^*_{\ell}
\varphi^*_{\ell+m} \varphi^{\vphantom*}_{\ell+n} 
\varphi^{\vphantom*}_{\ell+m-n}, \\
{\cal L}_1^{\rm even} & = & \sum_{\sigma,\sigma_1,\sigma_2=0}^{q-1} \!\!\!
\gamma^{\alpha \beta}_{\sigma_1\sigma_2} \varphi^*_{\alpha\sigma}
\varphi^{*}_{\beta,\sigma+\sigma_1} \varphi^{\vphantom*}_{\alpha,\sigma+\sigma_2} 
\varphi^{\vphantom*}_{\beta,\sigma+\sigma_1-\sigma_2}, \nonumber
\end{eqnarray}  
for odd and even $q$, respectively, where $\gamma_{mn}$ and
$\gamma_{\sigma\sigma'}^{\alpha\beta}$ are arbitrary at this stage.

Imposing additional restrictions on the coefficients in the above 
Lagrangians from invariance under rotations and reflections and taking
into account hermiticity and permutation symmetries, one may obtain a
set of conditions on the $\gamma$ coefficients required to preserve the
full triangular lattice symmetry.  These conditions are given in
Eqs.~(\ref{eq:17},\ref{eq:18}).  For specific $f$, they can readily be
solved to derive explicit forms for the Lagrangian.  
We will give these explicitly for $f=1/2$ and $1/3$ below. 

Interestingly, it is possible to make at least one general observation
concerning the symmetry properties of ${\cal L}_1^{\rm even}$.  All
terms in the continuum Lagrangian possess a global vortex $U(1)$ symmetry,
\begin{equation}
\label{eq:19}
\varphi_{\alpha\sigma} \rightarrow \varphi_{\alpha\sigma} e^{i \theta}.
\end{equation}
That is just a consequence of gauge invariance, expressing the
conservation of the bosonic current.  However, it is clear from
Eq.~(\ref{eq:16}) that the quartic Lagrangian ${\cal L}_1^{\rm even}$
possesses (at least) another, ``staggered'' U(1) symmetry:
\begin{equation}
\label{eq:20}
\varphi_{\alpha\sigma} \rightarrow \varphi_{\alpha\sigma} 
e^{i \alpha \theta}.
\end{equation}
This emergent $U(1)$ symmetry of the vortex theory implies that there
are (at least) two conserved dual charges, that can be labelled by the
index $\alpha$.  As discussed in Ref.~\onlinecite{Balents04}, this is
linked to the appearance of fractionally-charged bosonic excitations at
the critical point.  We will elaborate on the nature of these
excitations in Sec.~\ref{defects}.

\section{(Dual) Mean field theory}
\label{mft}

In this section we will discuss a mean field analysis of the vortex
theory, focusing on the nature of the ordered Mott insulating states
that occur.  We first present some general aspects of how spatial order
parameters are constructed in the vortex formalism, and give some
physical picture of how to think of the different Mott states.  The
remaining two subsections describe the specific phase diagram in the
case $f=1/3$ and the much more complicated and more interesting case
$f=1/2$.

\subsection{Order parameters and Mott states}

General arguments\cite{Oshikawa,Hastings} and physical reasoning seem to
imply that, barring exotic situations such as {\sl phases} with
``topological order'',\cite{Wen} Mott insulating states occuring for
non-integral $f$ must break space group symmetries (and in particular
translational symmetry).  Such space group symmetry breaking is measured
by spatial order parameters, the simplest of which (sufficient for our
purposes) describe non-uniformity of the boson density (beyond that
which is imposed by the underlying triangular substrate).

To visualize ordering patterns in the insulating phases we will find
it convenient to introduce a general ``density'' function
$\rho({\bf r})$, where ${\bf r}$ is a {\sl continuous} real-space
coordinate with ${\bf r}=0$ taken to coincide with a honeycomb lattice
site of type ``1'' (Fig.~\ref{hexlat}).  We construct $\rho({\bf r})$
to have the property that it transforms like a scalar boson density
under all symmetry operations.  It will be convenient to plot
$\rho({\bf r})$ to graphically illustrate the symmetry of the
non-uniform states that emerge in the theory.  Writing ${\bf r}=r_1
{\bf a}_1 + r_2{\bf a}_2$, one can actually construct such a function
quite generally for odd values of $q$:
\begin{equation}
\label{eq:26}
\varrho({\bf r}) = \sum_{m,n} \varrho_{m n} \omega^{m r_1 + n r_2},
\end{equation}
where the Fourier components $\varrho_{m n}$ serve as order parameters
for different ordered states, and are given by:
\begin{equation}
\label{eq:27}
\varrho_{m n} = S(m,n)\omega^{-mn/2 + (n-m)/6} \sum_{\ell=0}^{q-1} 
\omega^{-m\ell} \varphi^*_{\ell} \varphi_{\ell+n}.
\end{equation}
$S(m,n)$ here is a scalar form factor, that can not be determined from 
symmetry considerations, but should be chosen to depend only upon the
magnitude of the wavevector $m {\bf b}_1 + n {\bf b}_2$. 
A convenient and simple choice is the Lorentzian,
\begin{equation}
\label{eq:28}
S(m,n) = \frac{1}{1 + m^2 + (m+2 n)^2/3},
\end{equation}
which we use only for plotting purposes.  It is easy to check that
$\varrho_{mn}$ indeed transform like Fourier components of density:
\begin{eqnarray}
\label{eq:29}
&&T_1 : \varrho_{mn} \rightarrow \omega^{-m} \varrho_{mn}, \nonumber \\
&&T_2 : \varrho_{mn} \rightarrow \omega^{-n} \varrho_{mn}, \nonumber \\
&&R_{2\pi/3} : \varrho_{mn} \rightarrow \varrho_{n,-m-n}, \nonumber \\
&&I_{d_1} : \varrho_{mn} \rightarrow \varrho_{-n,-m}, \nonumber \\
&&I_{d_2} :\varrho_{mn} \rightarrow \omega^{(n-m)/3} \varrho_{nm}.
\end{eqnarray}

A similar function can be constructed for $q$ even.  When $f=1/q$, it takes 
the form
\begin{equation}
\label{eq:38}
\varrho({\bf r}) = \sum_{m,n} \left[\varrho_{mn}^{\alpha} + 
\tilde \varrho_{mn}^{\alpha} e^{2 \pi i \alpha (r_1 + r_2)/3 q}\right] 
\omega^{m r_1 + n r_2},
\end{equation}
where the density wave amplitudes are given by
\begin{eqnarray}
\label{eq:39}
&&\varrho^{\alpha}_{mn} = S(m,n) e^{\pi i \alpha (n-m)/3 q} 
\omega^{-mn/2 + (n-m)/6} \nonumber \\
&\times&\sum_{\ell=0}^{q-1} \omega^{-m\ell} \varphi^{\alpha *}_{\ell} 
\varphi^{\alpha}_{\ell+n}, \nonumber \\
&&\tilde \varrho^{\alpha}_{mn} = 
\tilde S(m,n) e^{i[\eta_2(-\alpha) - \eta_2(\alpha)]/2} 
\omega^{-mn/2 + (n-m)/6} \nonumber \\
&\times&\sum_{\ell=0}^{q-1} \omega^{-m\ell} \varphi^{-\alpha *}_{\ell} 
\varphi^{\alpha}_{\ell+n}.
\end{eqnarray}
Here the amplitude $\tilde S(m,n)$ has been taken in a simple form
consistent with rotational symmetry:
\begin{equation}
\label{eq:40}
\tilde S(m,n) = \frac{1}{1 + (m+\alpha/3)^2 + (m+2 n+\alpha)^2/3}.
\end{equation}
Eqs.~(\ref{eq:39}) are written in terms of $\eta_2(\alpha)$, which
also enters the PSG in general for even $q$, see
Eqs.~(\ref{eq:12},\ref{eq:13}) of
Appendix~\ref{sec:psg-momentum-space}.  It is hard to determine for
general $q$.  However, for the case we will focus upon, $f=1/2$, one
has
\begin{equation}
  \label{eq:eta2f12}
 \eta_2(\alpha)= -\frac{\pi\alpha}{12} \qquad \mbox{for $f=1/2$}.  
\end{equation}

We will present plots of $\rho({\bf r})$ for various mean field (and
beyond, in the following section!) Mott insulating states.

These images, and their deconstruction into the Fourier amplitudes
($\varrho_{mn}$ etc.), characterize the broken spatial symmetry of the
Mott states.  They do not, however, directly give a physical picture
for the ground state itself.  Of course, for a general interacting
boson model (and certainly in our approach where we do not specify the
microscopic Hamiltonian), we cannot hope to write down the ground
state wavefunction.  Moreover, this has far too much information.
What is conceptually useful is to understand how to write down a
simple wavefunction appropriate to a Mott insulator with the same
symmetries as predicted by the phenomenological theory.  By an
appropriate wavefunction, we mean one explicitly with the correct
boson filling, with no long-range correlations beyond that
of the Mott state, and consistent with incompressibility.  We consider
a satisfactory form to be a product state,
\begin{equation}
  \label{eq:wf}
  |\Psi\rangle = \prod_{\Box} |\psi(\Box)\rangle.
\end{equation}
The meaning of Eq.~(\ref{eq:wf}) is as follows.  We divide the lattice
into a set of non-overlapping identical clusters of sites -- unit
cells of the Mott state -- labeled by the $\Box$.  The $\prod$
indicates a direct product over states defined on the Hilbert space of
each box, with the same state $|\psi(\Box)\rangle$ chosen on each box.
The state $|\psi(\Box)\rangle$ must be an eigenstate of the total
boson number on the given cluster, for this wavefunction to represent
a Mott insulator, and this number should be chosen to match the
filling, i.e. equal to $f$ times the number of sites in the cluster.
Clearly this state has only local (within a cluster) charge
fluctuations, consistent with incompressibility.

Of course in most cases many such wavefunctions can be constructed for
a Mott state of a given symmetry, and moreover the true ground state
wavefunction for a realistic hamiltonian will not have the direct
product form.  However, the above general construction can serve the
purpose of providing an example of a state with the same symmetry
properties as predicted by the phenomenological theory, and is
expected to be in a sense (which we do not attempt to define
precisely) adiabatically connected to all ground states in the Mott
phase.  To our knowledge, all well-established examples of bose Mott
insulating ground states in the theoretical literature (e.g. on the
square lattice, checkerboard, stripe, and VBS states) have such a
construction.  We therefore view the existence of such a wavefunction
as a sort of consistency check on our results, and give examples in
the following subsections.  When the meaning is obvious, we give only
a brief physical description of the state, with the understanding that
one should keep the block product form, Eq.~(\ref{eq:wf}), in mind.

\subsection{$f=1/3$}
\label{q3}

At $1/3$-filling, the Mott state is not ``frustrated'' in the
colloquial sense.  The analysis below reveals that this case is
extremely analogous to the superfluid-Mott transition on the square
lattice at half-filling, and in particular provides another example of
a possible deconfined quantum critical point of the type discussed in
Ref.~\onlinecite{dcprefs}.

Solving Eq.(\ref{eq:18}) at $f=1/3$, we obtain the following form of the
quartic potential in the continuum Lagrangian density:
\begin{eqnarray}
  \label{eq:21}
  &&{\cal L}_1 = u \left(|\varphi_0|^2 + |\varphi_1|^2 + |\varphi_2|^2\right)^2
  \nonumber \\
  &+&v\left\{|\varphi_0|^2 |\varphi_1|^2 + |\varphi_1|^2 |\varphi_2|^2 + 
    |\varphi_2|^2 |\varphi_0|^2 + 2 {\textrm Re} \left[e^{-2 \pi i/3} 
    \right.\right. \nonumber \\ 
  &\times&\left.\left.\left(\varphi_0^{* 2}\varphi_1 \varphi_2 + 
        \varphi_2^{* 2} \varphi_0 
        \varphi_1 + \varphi_1^{* 2} \varphi_2 \varphi_0 + c.c.\right)
    \right]\right\}. 
\end{eqnarray}
As discussed in Ref.~\onlinecite{Balents04}, it is convenient to
transform to a different set of variables that realize a ``permutative
representation'' of the PSG.  As in the square lattice case, such
representations exist only for special fillings.  In particular, it is
easy to show that a permutative representation does not exist at
$f=1/2$, in contrast to the square lattice case.  At $f=1/3$ however, it
does exist and is given by:
\begin{eqnarray}
\label{eq:22}
\xi_0 = \frac{1}{\sqrt{3}} \left(\varphi_0 + e^{-2 \pi i /3} \varphi_1 + 
  \varphi_2 \right), \nonumber \\
\xi_1 = \frac{1}{\sqrt{3}} \left(e^{-2 \pi i/3} \varphi_0 + \varphi_1 + 
  \varphi_2 \right), \nonumber \\
\xi_2 = \frac{1}{\sqrt{3}} \left(\varphi_0 + \varphi_1 + 
  e^{-2 \pi i /3} \varphi_2 \right).
\end{eqnarray}
The physical meaning of the three vortex quantum numbers in the
permutative representation of the PSG is that they represent three
conserved (as shown below) dual charges (vorticities).  The quantum
numbers that are dual to these three vorticities are three real
fractional (1/3 of the boson charge) charges.  That is, a dual ``vortex'' in
which the phase of any one of the $\xi_\ell$ fields winds by $\pm 2\pi$
at infinity carries a localized charge (boson number) of $\pm 1/3$ (see
Ref.~\onlinecite{Balents04} for a simple derivation of this result in a
more general context).

The representation of the PSG realized by the $\xi_\ell$ fields is
permutative in that each symmetry operation is realized as the
composition of a permutation
of the $\xi_\ell$ fields and a simple phase rotation.  In particular,
\begin{eqnarray}
\label{eq:23}
&&T_1 : \xi_{\ell} \rightarrow \xi_{\ell+1}, \nonumber \\
&&T_2 : \xi_{\ell} \rightarrow e^{2 \pi (\ell+1) i/3} \xi_{\ell+1}, 
\nonumber \\
&&R_{2 \pi/3} : \xi_0 \rightarrow e^{\pi i/6} \xi_2,\,\,
\xi_1 \rightarrow -i \xi_0,\,\, 
\xi_2 \rightarrow  e^{\pi i/6} \xi_1, \nonumber \\
&&I_{d_1} : \xi_0 \rightarrow e^{-\pi i /6} \xi^*_2,\,\,
\xi_1 \rightarrow -e^{\pi i/6} \xi^*_1,\,\,
\xi_2 \rightarrow e^{-\pi i /6}\xi^*_0,\nonumber\\
&&I_{d_2} : \xi_0 \rightarrow e^{-\pi i /6} \xi^*_0,\,\,
\xi_1 \rightarrow -e^{\pi i/6} \xi^*_1,\,\,
\xi_2 \rightarrow e^{-\pi i /6}\xi^*_2. \nonumber \\
\end{eqnarray}
The quartic potential simplifies greatly in these variables:
\begin{eqnarray}
\label{eq:24}
&&{\cal L}_1 = u \left(|\xi_0|^2 + |\xi_1|^2 + |\xi_2|^2\right)^2 \nonumber \\
&+& v \left(|\xi_0|^2 |\xi_1|^2 + |\xi_1|^2 |\xi_2|^2 + |\xi_2|^2 
|\xi_0|^2 \right)
\end{eqnarray}
At the quartic level, it is immediately apparent that the microscopic
overall $U(1)$ gauge symmetry required by the
vortex non-locality has been elevated to a $U(1)^3$ symmetry under
independent rotations of all three $\xi_\ell$ fields.  Of this, only the
group of equal rotations of all fields is gauge, leaving an additional
$U(1)^3/U(1) = U(1)\times U(1)$ global (not gauge) symmetry of the dual
theory.  This emergent symmetry is broken at $6^{\rm th}$ order by the term
\begin{equation}
\label{eq:25}
{\cal L}_2 = w \left[ (\xi^*_0 \xi_1)^3 + (\xi^*_1 \xi_2)^3 + 
(\xi^*_2 \xi_0)^3 + c.c.\right] .
\end{equation}
The structure of the theory is thus rather similar to the vortex theory
at $f=1/2$ on the square lattice.\cite{Lannert01,Balents04} It provides
another example of deconfined criticality, if, as seems likely, the
mean-field irrelevance of the higher-order term in ${\cal L}_2$ remains
valid with fluctuations in $2+1$ dimensions, for some sign of $v$. The
mean field phase diagram of the vortex theory can now be easily found
analytically.
\begin{figure}[t]
\includegraphics[width=6cm]{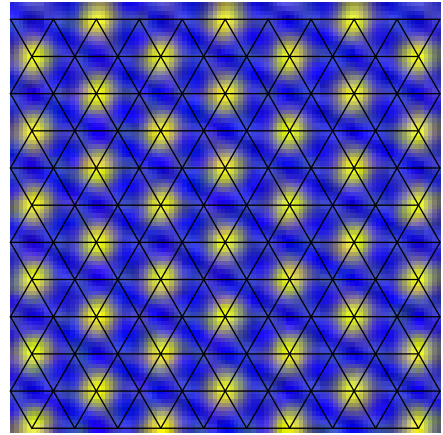}
\caption{Charge density pattern at $1/3$-filling for $v > 0$.} 
\label{q3vpos}
\end{figure}

\begin{figure}[t]
\includegraphics[width=6cm]{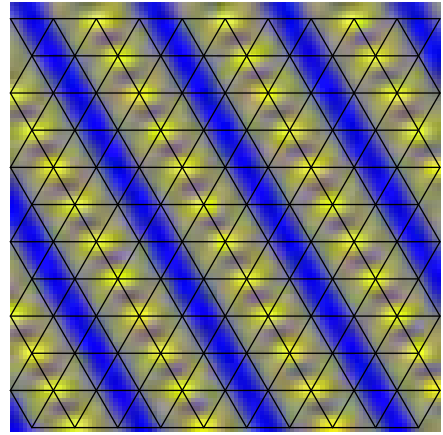}
\caption{Charge density pattern at $1/3$-filling for $v < 0$ and $w < 0$.} 
\label{q3vnegwneg}
\end{figure}
Let us now proceed with the mean field theory for $f=1/3$.  For $s <
0$ and $ |w| < |v|$ one obtains 3 distinct insulating phases:

\vspace{0.1in}\noindent\underline{\bf I. $\boldsymbol{v > 0}$:}\\
The energy is minimized if only one of the 3 vortex flavors condenses.
This state is then clearly 3-fold degenerate and breaks all symmetries
except reflection with respect to $d_2$ axis.  In Fig.\ref{q3vpos}
this state is visualized explicitly by plotting the corresponding
density wave order parameter.  This state is the simplest CDW state at
$1/3$-filling, with an example wavefunction consisting of a boson
number eigenstate on each site.  Microscopically this phase is the
natural Mott insulating state in a model in which the Mott transition
is driven by strong nearest-neighbor repulsive interactions, like the
XXZ model.  We do not expect the superfluid-Mott transition to this
state is likely to be continuous or deconfined when fluctuations are
taken into account in $2+1$ dimensions.

\vspace{0.1in} \noindent \underline{\bf II. $\boldsymbol{v < 0}$:}\\
It is energetically favorable to condense all 3 vortex flavors, so
that all vortex fields have equal magnitude.  There are then 2
different phases, depending on the sign of $w$.  These states are
those expected to be connected by a deconfined quantum critical point
to the superfluid state.
\begin{enumerate}
\item $w < 0$.\\
  Writing $\xi_{\ell} \sim |\xi| e^{i\theta_{\ell}}$, the minimum is
  achieved when:
\begin{equation}
\label{eq:30}
\theta_1-\theta_0 = 2\pi m/3, \theta_2-\theta_0 = 2 \pi n /3, 
\end{equation}
for $m,n=0,1,2$.
\begin{figure}[t]
\includegraphics[width=6cm]{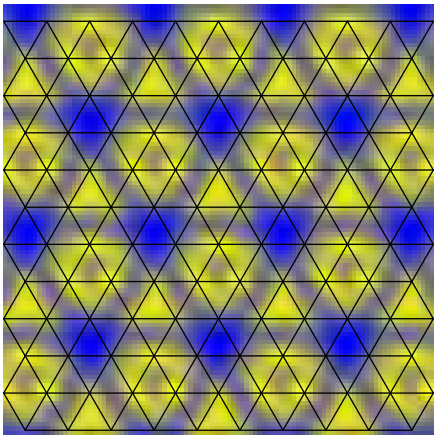}
\caption{Charge density pattern at $1/3$-filling for $v < 0$ and $w > 0$.} 
\label{q3vnegwpos}
\end{figure}
This state is thus 9-fold degenerate and corresponds to period-3
site-centered stripes, running in ${\bf a}_1, {\bf a}_2$ and ${\bf
  a}_1 + {\bf a}_2$ directions, see Fig.\ref{q3vnegwneg}.  One may
construct a wavefunction for this state by e.g. taking linear 3-site
clusters at a $60^\circ$ angle to the stripe (e.g. horizontal in
Fig.~\ref{q3vnegwneg}), and putting a boson at the center of the
cluster, leaving the other sites empty.  
\item $w > 0$.\\
One can readily verify that the ground state is 18-fold degenerate, the
distinct solutions being obtained from
\begin{equation}
  \label{eq:18fold}
  \theta_1-\theta_0=\theta_2-\theta_0=2\pi/9,
\end{equation}
by applying translations and the reflection $I_{d_2}$.  
The corresponding characteristic density pattern is shown in
Fig.\ref{q3vnegwpos}. It is adiabatically connected to a ``bubble''
phase or crystalline state in which one boson is placed on each site
of each elementary triangle (3 bosons total) on a $3\times 3$
triangular superlattice.  The 18-fold degeneracy results from 9 states
obtained by translating this pattern, and another 9 states obtained by
choosing say down-pointing instead of up-pointing triangles.
\end{enumerate}

\subsection{$f=1/2$}
\label{sec:f=12}

\subsubsection{Action and symmetries}
\label{sec:action-symmetries}

We now turn to the more complicated and interesting case
of $f=1/2$.  It is straightforward to show (by a simple generalization of the
argument used in Ref. \onlinecite{Balents04} to prove the absence of
a permutative representation in the $f=1/3$ case on the square
lattice),  that for this case there is no permutative representation
of the PSG.  Unlike in the square lattice case, there are also,
surprisingly, {\sl more} low-energy vortex modes for $f=1/2$ case than for
$f=1/3$.  The emergent low-energy symmetry amongst these modes is,
moreover, nonabelian, with an $SU(2)$ ``pseudo-spin'' subgroup that we
will uncover below.  This structure has interesting consequences for
the phases and excitations that will be explored here and in the
following section in some detail.
 
The quartic potential can be found by solving Eq.(\ref{eq:17}). 
It turns out to have the following form:
\begin{eqnarray}
\label{eq:32}
&&{\cal L}_1 = u (\varphi^*_{\alpha\sigma}
\varphi^{\vphantom*}_{\alpha\sigma})^2 + v (\varphi^*_{+\sigma}
\varphi^{\vphantom*}_{+\sigma})(\varphi^*_{-\sigma'}
\varphi^{\vphantom*}_{-\sigma'})  \\
&&+w\left(|\varphi_{-0}|^2 |\varphi_{+0}|^2 + \varphi^*_{-1} \varphi^*_{+1} \varphi_{-0} \varphi_{+0} +
(0\leftrightarrow 1) \right). \nonumber
\end{eqnarray}
To make the symmetries of the Lagrangian more transparent, it is useful 
to rewrite it as follows. First we define $z_{\alpha\sigma}$ pseudospinor
variables,
\begin{eqnarray}
\label{eq:33}
\varphi_{-\sigma} & = & -\epsilon_{\sigma\sigma'} z_{-\sigma'},
\nonumber \\
\varphi_{+\sigma} & = & z_{\sigma}, 
\end{eqnarray} 
where $\epsilon_{\sigma\sigma'}$ is the antisymmetric tensor with
$\epsilon_{01}=-\epsilon_{10}=1$.  The quartic action has $SU(2)$
symmetry under rotations of the $\sigma$ index of $z_{\alpha\sigma}$.
This is made manifest by introducing the pseudospin vector
\begin{equation}
\label{eq:34}
S^a_{\alpha} = z^*_{\alpha\sigma} \tau^a_{\sigma \sigma'} 
z^{\vphantom*}_{\alpha\sigma'}, 
\end{equation}
where $\tau^a,\, a=x,y,z$ are the Pauli matrices, and summation on
the repeated $\sigma,\sigma'$ indices is implied.  The transformation
properties of the ${\bf S}_\pm$ vectors are particularly simple, and
given in Appendix~\ref{sec:pseud-transf}.  In terms of these
pseudospin variables the quartic potential becomes, after a trivial
redefinition of $v$ and $w$ couplings:
\begin{equation}
\label{eq:35}
{\cal L}_1 = u \left(S_+ + S_-\right)^2 + v \,S_+ S_- + 
w_1 \,{\bf S}_+ \cdot {\bf S}_- ,
\end{equation}
where $S_\alpha = |{\bf S}_\alpha| = \sqrt{z_{\alpha\sigma}^*
  z_{\alpha\sigma}^{\vphantom*}}$ (sum on $\sigma$ implied).  It is
now clear that the quartic potential has, in addition to the
microscopic gauge $U(1)$ symmetry, an $SU(2) \times U(1) \times Z_2$
invariance.  The $SU(2)$ symmetry is manifest in Eq.(\ref{eq:35}), the
extra $U(1)$ symmetry is the ``staggered'' $U(1)$, already mentioned
above, see Eq.(\ref{eq:20}), also manifest since $S^a_\alpha$ are
independent of the staggered $U(1)$ phase.  The $Z_2$ symmetry is the
interchange ${\bf S}_+\leftrightarrow {\bf S}_-$ (particle-hole
symmetry $C$ in fact requires this invariance up to a sign, though
Eq.~(\ref{eq:35}) is obtained without using $C$).  The pseudospin
variables are directly related to the $\varrho^{\alpha}_{mn}$ density
components:
\begin{eqnarray}
\label{eq:41}
&&\varrho^{\alpha}_{00} = S_{\alpha}, \nonumber \\
&&\varrho^{\alpha}_{01} = \alpha e^{\pi i (1 + \alpha)/6} S_{\alpha}^x, 
\nonumber \\
&&\varrho^{\alpha}_{10} = \alpha e^{-\pi i (1 + \alpha)/6} 
S_{\alpha}^z, \nonumber \\
&&\varrho^{\alpha}_{11} = S_{\alpha}^y.
\end{eqnarray}
The physical import of the $SU(2)$ symmetry of Eq.~(\ref{eq:35}) is
now clear: all CDW states, related to each other by arbitrary
rotations in the space of the three Fourier components of the density,
are degenerate at this order.  This degeneracy will be weakly broken,
of course, by higher order terms in the action.  As discussed in the
introduction, the large emergent symmetry is thereby connected with
geometrical charge frustration at $f=1/2$.

\subsubsection{Order parameters}
\label{sec:order-parameters}
  
The pseudospin vectors ${\bf S}_\pm$ serve as gauge-invariant order
parameters to characterize the breaking of the $SU(2)$ symmetry.  It
is instructive to construct two other such order parameters.
The emergent $U(1)$ symmetry is best characterized by a complex order
parameter, $\psi$, defined by
\begin{equation}
  \label{eq:psidef}
  \psi = e^{i\pi/4} z_{+\sigma}^*  z_{-\sigma}^{\vphantom*},
\end{equation}
where we have included the $\pi/4$ phase factor for later
convenience.  One may also define an Ising order paramer $\Phi$, with
\begin{equation}
  \label{eq:Ising}
  \Phi = |z_+|^2 - |z_-|^2.
\end{equation}
Non-zero $\langle \Phi \rangle$ implies $I_{d_2}$ and $C$ are broken.
The physical meaning of non-vanishing ${\bf S}_\pm$ and $\psi$ will be
elucidated in detail in the following.

Though the terms which break these symmetries are small near the Mott QCP
(hopefully irrelevant there), they are important at sufficiently low
energy.  We must therefore consider those higher order terms in the
action which are required to reduce the $SU(2) \times U(1)$ symmetry
to only what is required by the PSG. Higher order terms which do not
reduce this symmetry need not be considered.

First consider the $SU(2)$ pseudospin symmetry.  In the absence
of microscopic particle-hole invariance, it is broken at $6^{\rm th}$
order by a term of the form $S_+^x S_+^y S_+^z + S_-^x S_-^y S_-^z$.
We will, however, for concreteness focus on the case relevant to the
XXZ model and other pairwise interacting boson lattice models, in
which particle-hole symmetry is an invariance of the theory.  In this
case, the $SU(2)$ symmetry is broken only at the 8th order by a
``cubic anisotropy'' term,
\begin{equation}
\label{eq:36}
{\cal L}_2 = w_2 \sum_\alpha \left[\left(S^x_{\alpha}\right)^4 +
\left(S^y_{\alpha}\right)^4 + \left(S^z_{\alpha}\right)^4 \right].
\end{equation}
The ``staggered'' $U(1)$ symmetry is more persistent and is broken at
the 12th order.  The simplest term at this order that breaks the
``staggered'' $U(1)$ symmetry is:
\begin{eqnarray}
\label{eq:37}
{\cal L}_3 = - w_3\, {\rm Re} \left( \psi^6 \right).
\end{eqnarray}

\begin{figure}[t]
\includegraphics[width=6cm]{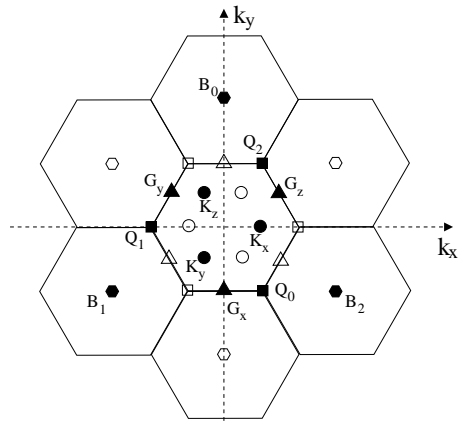}
\caption{Ordering wavevectors in the reciprocal lattice.  Filled
  symbols are the wavevectors referred to in the text.  Open symbols
  indicate wavevectors belonging to the same star which are related to
  the former sets by complex conjugation.} 
\label{hexbz}
\end{figure}        

It is useful to reorganize various terms in the density expansion of
Eq.~(\ref{eq:38}) to understand in more detail the nature of the
different order parameters.  The wavevectors
referred to in the following are labelled in Fig.~\ref{hexbz}.  

Consider first  Ising order.  Non-zero $\langle \Phi\rangle$
implies only that the reflection $I_{d_2}$ and particle-hole symmetry
$C$ are broken.  Thus it corresponds only to a modulation of the
density {\sl within} the primitive unit cell of the triangular
lattice.  The corresponding density modulation therefore occurs entirely
at reciprocal lattice vectors.  The smallest set of reciprocal lattice
wavevectors that can describe the modulation are ${\bf B}_0=(0,2\pi)$,
${\bf B}_0=(2\pi,-2\pi)$, ${\bf B}_2=(-2\pi,0)$.  This density modulation is
\begin{equation}
  \label{eq:isingdens}
  \rho_\Phi({\bf r}) = \Phi \sum_{n=0,1,2} \cos ({\bf B}_n \cdot {\bf r} +
  \frac{5\pi}{6}) .
\end{equation}
While $\rho_\Phi=0$ on triangular (direct) lattice sites (as it must), it
alternates sign on sites of the dual honeycomb lattice, i.e. takes
opposite signs on centers of triangles of the direct lattice.

Now let us turn to pseudospin ordering.  The existence of non-vanishing
$\langle {\bf S}_\pm\rangle$ implies charge ordering at the
wavevectors ${\bf G}_x=(0,-\pi)$, ${\bf
  G}_y=(-\pi,\pi)$, ${\bf G}_z=(\pi,0)$, which
lie at the centers of the zone edges.  In particular, the associated
density modulations take the form
\begin{equation}
  \label{eq:psvecdens}
  \rho_S({\bf r}) = \sum_{a=x,y,z} {\rm Re}\, \left[ (S_+^a e^{-i\pi/3}
    - S_-^a ) e^{i {\bf G}_a\cdot{\bf r}}\right],
\end{equation}
neglecting higher harmonics which do not change the symmetry of $\rho_S({\bf
  r})$.  

Next consider the XY order parameter $\psi$.  A state with $\langle
\psi\rangle$ exhibits a three-sublattice structure, characterized by
the zone boundary wavevectors ${\bf Q}_0=2\pi(\frac{2}{3},-\frac{1}{3})$, ${\bf
  Q}_1=2\pi(-\frac{1}{3},\frac{2}{3})$, ${\bf
  Q}_2=2\pi(-\frac{1}{3},-\frac{1}{3})$: 
\begin{equation}
  \label{eq:xydens}
  \rho_{\psi}({\bf r}) =  {\rm Re}\, \left[ \psi e^{-i\pi/6} \sum_{n=0,1,2}\, 
    e^{i 2\pi n/3} e^{i{\bf Q}_n \cdot {\bf r}} \right].
\end{equation}
In fact, these three wavevectors differ only by reciprocal lattice
vectors.  From this, it is straightforward to show that
$\rho_\psi({\bf r})$ vanishes on dual lattice sites, so that all
triangular plaquettes of the direct lattice are equivalent up to
rotations in an XY ordered state.

Finally, simultaneous breaking of pseudospin and XY symmetry is
characterized by the ``composite'' order parameter ${\bf d}$, a
complex vector, defined as
\begin{equation}
  \label{eq:dvec1}
  {\bf d} = z_{+\sigma}^* {\boldsymbol \tau}_{\sigma\sigma'}
  z_{-\sigma'}^{\vphantom*}.
\end{equation}
When $\langle {\bf d}\rangle\neq 0$, density modulations appear at the
wavevectors ${\bf K}_x = 2\pi(\frac{1}{3},-\frac{1}{6})$, ${\bf K}_y =
2\pi(-\frac{1}{6},-\frac{1}{6})$, ${\bf K}_z =
2\pi(-\frac{1}{6},\frac{1}{3})$, which lie {\sl within} the zone.  The
corresponding density is
\begin{equation}
  \label{eq:rhod}
  \rho_{d}({\bf r}) = {\rm Re}\, \left[ e^{3\pi i/4} \sum_{a=0,1,2}
    d_a e^{2\pi (a-1)i/3} e^{i{\bf K}_a\cdot{\bf r}} \right],
\end{equation}
where we have identified $a=x,y,z$ with $a=0,1,2$ respectively.

\subsubsection{Mean field phases}
\label{sec:mean-field-phases}

\begin{figure}[t]
\includegraphics[width=6cm]{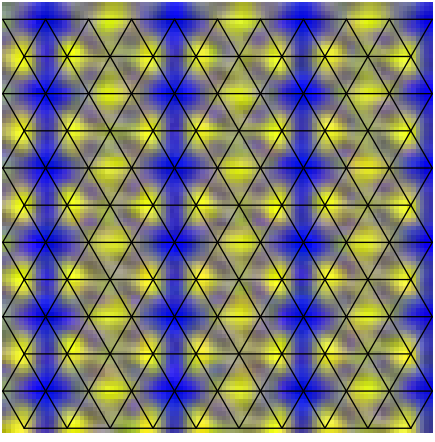}
\caption{Charge density pattern at $1/2$-filling for $v < 0$, $w_1 < 0$, 
$w_2 < 0$ and $w_3 < 0$.} 
\label{q2vnegwnegw1neg}
\end{figure}     

The mean field phase diagram of ${\cal L}_0 + {\cal L}_1 + {\cal L}_2 + 
{\cal L}_3$ can be easily obtained analytically. 
One finds 10 different phases: 2 for $v > 0$ and 8 for $v < 0$. 
We will not attempt to be exhaustive in describing these states.   We
will, however, discuss {\sl some} of the phases in detail, and go into
some general aspects of the 8 cases with $v<0$ in
Sec.~\ref{defects}.  Minimizing the mean-field energy functional,
$v<0$ implies that both $\alpha=\pm$ vortex pseudospinors are
condensed with equal amplitude, so $S_+=S_-$.    
The sign of $w_1$ determines the relative pseudospin orientation. For
$w_1 < 0$, they are parallel, i.e. ${\bf S}_+={\bf S}_- \equiv {\bf S}$.
In terms of the vortex variables this condition is most generally 
solved by
\begin{equation}
  \label{eq:43}
  z_{\pm\sigma} = z_{\sigma} e^{\pm i\theta/2},
\end{equation}
where ${\bf S}=z_\sigma^* {\boldsymbol \tau}_{\sigma\sigma'}
z_{\sigma'}^{\vphantom*}$.  For $w_1>0$, the two pseudospin vectors
are antiparallel, ${\bf S}_+=-{\bf S}_- \equiv {\bf S}$, which implies
\begin{eqnarray}
  \label{eq:44}
  z_{+\sigma}&=&z_{\sigma} e^{i \theta/2}, \nonumber \\
  z_{-\sigma}&=&\epsilon^{\vphantom*}_{\sigma \sigma'} z^*_{\sigma'} 
  e^{i \theta/2}.
\end{eqnarray}
We note that the dual gauge symmetry acts differently in the two
cases.  For parallel pseudospins, a gauge transformation takes
$z_\sigma \rightarrow e^{i\chi}z_\sigma$, while for antiparallel
pseudospins, instead $\theta\rightarrow \theta + 2\chi$.

The remaining terms, $w_2$ and $w_3$, fix the remaining non-gauge
symmetries.  The sign of $w_2$ chooses the easy axes of the pseudospin
vector, along $(100)$ and symmetry-related axes for $w_2<0$, and along
$(111)$ and related axes for $w_2>0$.  The above conditions leave only
the relative phase between the spinors $z_\pm$, corresponding to the
``staggered'' $U(1)$ symmetry of the corresponding terms in the vortex
Lagrangian.

This relative phase is fixed at $12^{\rm th}$ order in $z_{\pm\sigma}$,
for instance for $w_1<0$ by the ${\cal L}_3$ term (for $w_1>0$, a more
complicated $12^{\rm th}$ order term must be included as the $w_3$
interaction vanishes in that case).  Depending on the sign of $w_3$, the
energy minimum is achieved when $\sin (6 \theta) = \pm 1$, i.e.
\begin{equation}    
\label{eq:52}
\theta = \frac{\pi}{12} + \frac{\pi n}{3}, \,\, n=0,\ldots,5,
\end{equation} 
or
\begin{equation}    
\label{eq:53}
\theta = \frac{\pi}{4} + \frac{\pi n}{3}, \,\, n=0,\ldots,5,
\end{equation} 
One of the states, corresponding to $\theta = \pi/12$ and
\begin{eqnarray}
\label{eq:54}
z_{+0} &=&z_{+1} = \frac{e^{i \theta/2}}{\sqrt{2}}, \nonumber \\
z_{-0}&=&z_{-1} = \frac{e^{-i\theta/2}} {\sqrt{2}},
\end{eqnarray} 
is shown in Fig.\ref{q2vnegwnegw1neg}.
We will elucidate the physics of
this particular pair of states in some detail in the discussion section.
 
We can generally classify all the mean field states by their degeneracies and 
the corresponding unit cell sizes. The states with the pseudospin easy 
axis along (100) and parallel pseudospins (the state in 
Fig.\ref{q2vnegwnegw1neg} belongs to this group) are all 36-fold degenerate 
and have a 6-site unit cell. 
States with parallel pseudospins but with the easy axis along (111) are 
48-fold degenerate and have the largest, 12-site unit cell. 
In the case of antiparallel pseudospins, ground state degeneracies are the 
same, but the unit cell size of the 36-fold degenerate states doubles 
to 12 sites. 
States with the smallest unit cells are obtained when only one of the 
pseudospinors is condensed. 
In this case one obtains a 6-fold degenerate ground state and a 2-site unit 
cell when (100) is the easy axis, and an 8-fold degenerate state with a 
4-site unit cell when the easy axis is along the (111) direction.

For the phases of most interest ($v<0$) in which the pseudospin
vectors are either parallel or antiparallel, some of the density functions
associated to the order parameters in Sec.~\ref{sec:order-parameters}
can be simplified to an extent.  The only qualitative 
case is the pseudospin vector density $\rho_S$.
When ${\bf S}_+={\bf S}_-={\bf S}$, it reduces to   
\begin{equation}
  \label{eq:psvecdenspar}
  \rho_{S\parallel}({\bf r}) = \sum_{a=x,y,z} S^a \cos ({\bf
    G}_a\cdot{\bf r} - \frac{2\pi}{3}).
\end{equation}
Most interesting, when ${\bf S}_+=-{\bf S}_-={\bf S}$, one has instead
\begin{equation}
  \label{eq:psvecdensapar}
    \rho_{S,\not\parallel}({\bf r}) = \sum_{a=x,y,z} S^a \cos ({\bf
    G}_a\cdot{\bf r} - \frac{\pi}{6}).
\end{equation}
In this case, it is noteworthy that $\rho_{S,\not\parallel}$ vanishes
on all triangular lattice sites.  This is a consequence of the fact
that a configuration of antiparallel pseudospins preserves
particle-hole symmetry, so modulations can occur only in bond or
plaquette ``kinetic'' terms.  Thus ${\bf S}_+ - {\bf S}_-$ may be
considered a (particular) purely valence bond solid order parameter.

The composite order parameter ${\bf d}$ also simplifies once the
pseudospin order is determined.  For parallel pseudospins, one simply
has ${\bf d}={\bf S}$.  For antiparallel pseudospins, instead, one has
${\bf d}\times {\bf d}^* = 2 i{\bf S}$, which implies
\begin{eqnarray}
  \label{eq:dvecasab}
  &&  {\bf d} = {\bf e}_1 - i {\bf e}_2, \qquad  {\bf e}_1\times{\bf
    e}_2= {\bf S},
\end{eqnarray}
so that ${\bf e}_1,{\bf e}_2,{\bf S}$ form a right-handed orthogonal
frame in the O(3) spin space.  The angle of ${\bf e}_1$ in the XY
plane is arbitrary, and determined by (twice) the phase of $z_\sigma$.

\section{``Hard Spin'' Description: Beyond Mean-Field Theory}
\label{defects}

In the preceding section, we have followed a Landau-theory like
procedure (albeit with non-LGW vortex fields) in expanding the effective
action in a power series in the $\varphi_\ell$ fields, whose amplitude
is viewed as small in the vicinity of the Mott QCP.  In low-dimensional
statistical mechanics, it is often preferable to formulate the theory in
terms of ``hard spin'' variables, in which the amplitude of the order
parameter field(s) is fixed, and only the ``angular'' degrees of
freedom are free to fluctuate and vary in space and time.  Examples
include the Kosterlitz-Thouless theory of the XY phase transition, and
the non-linear sigma model formulation of $O(n)$ models.  The intuitive
rationale for such an approach is that, in low dimensions, fluctuations
suppress the ordering point of the transition well below the mean-field
point, so that substantial amplitude is already developed in the true
critical region.  Whatever the rationale, there are some advantages to
such a hard-spin approach.  Duality transformations generally apply to
hard-spin models.   Hard-spin variables are particularly appropriate to
describe the elementary excitations of ``ordered'' phases in which the
amplitude of the fields is on average large, and only the Goldstone-like
fluctuations of the orientation of these fields comprise low-energy
excitations.  Finally, a hard-spin formulation returned to the lattice
is fully-regularized, and can thereby address non-perturbative phenomena
in a controlled manner.

In this section, we will provide and analyze a hard-spin
formulation of the dual vortex action.  These allow us to identify the
excitations and their quantum numbers within the Mott phases.  Notably,
we find that the most interesting Mott states support two distinct kinds
of excitations.  First, there are $1/2$-charged (``spinon'' in the
spin-$1/2$ XXZ language) ``vortex'' excitations which are linearly
confined in pairs deep in the Mott state, but are only logarithmically
interacting up to a long ``confinement length'' near the superfluid-Mott
transition.  Second, there are additional unit charged ``skyrmion''
excitations which are everywhere deconfined in the Mott state.  These
are adiabatically connected to single boson vacancies/interstitials in
the Mott solid, but become topological as the Mott transition is
approached.  The hard-spin models also provide a firm ground on which to study
other phases, notably {\sl supersolids}, which do not occur within a
mean field treatment of the vortex field theory, but are extremely
natural in this approach.

\subsection{Formulation of hard spin model}
\label{sec:form-hard-spin}

To write down an appropriate hard-spin model, we imagine tuning $s<0$
to a point beyond the mean field Mott transition point.  At such a
point, the minimum action configurations have non-zero amplitude.  We
will assume the mean amplitude is determined by a balance between the
quadratic Lagrangian, Eq.~(\ref{eq:vfteven}) and the quartic terms in
${\cal L}_1$, Eq.~(\ref{eq:35}).  The minimum action saddle points of
the combination of these two terms are constant in space-time, but
allow for a continuous set of orientations in the field space.  We
focus in particular on $v<0$, so that the magnitude $S_+=S_-$ is
fixed.  We will focus primarily on the case
$w_1<0$, so that the saddle point has ${\bf S}_+={\bf S}_-$.  This is
the most interesting case,  because, as we shall see, the
recently-determined supersolid phase of the XXZ model can be
understood in this framework.  At the end of this section, we will
briefly summarize the results of similar analysis for $w_1>0$,
corresponding to anti-parallel pseudospins.

We further suppose that $s$ is negative enough that fluctuations in
the above conditions may be neglected, but that within these constraints
the $z_{\alpha\sigma}$ fields can vary spatially.  We will
therefore absorb any effects of the magnitude of the fields into
coefficients, and without loss of generality normalize to $S_+=S_-=1$.
Ultimately, the higher order terms will still be included, but can be
considered small perturbations.

The most general solution of ${\bf S}_+={\bf S}_- = {\bf S}$ constraint 
in terms of the vortex
variables is given by Eq.~(\ref{eq:43}).  We will therefore rewrite
the action in terms of the CP$^{1}$ field $z_{\sigma}$ and an XY
field $e^{i\theta/2}$.  Note that this solution
possesses a $Z_2$ gauge invariance under
\begin{equation}
  \label{eq:z2g}
  z_\sigma \rightarrow - z_\sigma, \qquad \theta \rightarrow \theta+2\pi.
\end{equation}
This is in addition to the physical symmetries of the model.  It is a
gauge invariance since it can be performed independently at each
space-time point without changing $z_{\alpha\sigma}$, and hence
physical quantities.

Inserting Eq.~(\ref{eq:43}) into the action, and regularizing it on
a space-time lattice, we obtain 
\begin{equation}
  \label{eq:Shs}
  {\cal L}_{\rm HS}^{\parallel} =- t_v e^{-i A_{i\mu}} z^*_{i \sigma} z_{i+\mu
      \sigma} \cos(\Delta_\mu \theta_i/2) +
    \frac{1}{2e^2}(\epsilon_{\mu \nu \lambda} \Delta_{\nu} A_{i \lambda})^2,
\end{equation}
with $z_{i\sigma}^* z_{i\sigma}^{\vphantom*}=1$ normalized on each
site $i$ of the space-time lattice.  Note that Eq.~(\ref{eq:Shs}) is
indeed invariant independently under Eq.~(\ref{eq:z2g}) at each point
$i$.  It is convenient to rewrite the first term in
Eq.~(\ref{eq:Shs}), making the Ising gauge symmetry explicit by
introducing an Ising gauge field $\sigma_{i\mu}$ which resides on the
link $(i,i+\mu)$:
\begin{eqnarray}
\label{eq:Lz2}
  {\cal L}_{\rm Z_2}^{\parallel} & = & - t_z \sigma_{i\mu} e^{-i
    A_{i\mu}} z^*_{i \sigma} z_{i+\mu \sigma} - t_\theta \sigma_{i\mu}
  \cos(\Delta_\mu 
  \theta_i/2) \nonumber \\
  && +  \frac{1}{2e^2}(\epsilon_{\mu \nu \lambda} \Delta_{\nu} A_{i
    \lambda})^2,
\end{eqnarray}
where we have introduced two independent parameters $t_z,t_\theta$.
One could add a plaquette interaction (line product of $\sigma_{i\mu}$
around plaquettes), which is a standard ``kinetic term'' for the $Z_2$
gauge field for further generality.  We are, however, most interested
in the limit in which it is absent.  In this case, one may sum over
$\sigma_{i\mu}$ on each link independently, to return to an action of
the form of Eq.~(\ref{eq:Shs}) (with additional higher-order terms).
We will not attempt, however, to constrain Eq.~(\ref{eq:Lz2}) to be
exactly equivalent to Eq.~(\ref{eq:Shs}).  Instead, since we are
anyway constructing a phenomenological theory, we regard the freedom
to vary $t_z,t_\theta$ independently as a means of capturing the
different possible tendencies due to fluctuation effects and details
of microscopic dynamics in different physical systems.

\subsection{Phase diagram for parallel pseudospins}
\label{sec:phase-diagr-parall}

Let us now discuss the phase diagram of Eq.~(\ref{eq:Lz2}).  For
$t_z,t_\theta \ll 1$, both $\theta_i$ and $z_{i\sigma}$ variables are
disordered, and the dual gauge field $A_{i\mu}$ is gapless.  This is
the superfluid phase.  For $t_\theta \sim t_z \gg 1$ large and of the
same order, we expect that both the CP$^1$ and XY variables are
ordered.  This is the Mott insulator, whose precise nature depends
upon the anisotropy terms we have neglected to write.  

\begin{figure}[t]
\includegraphics[width=4cm]{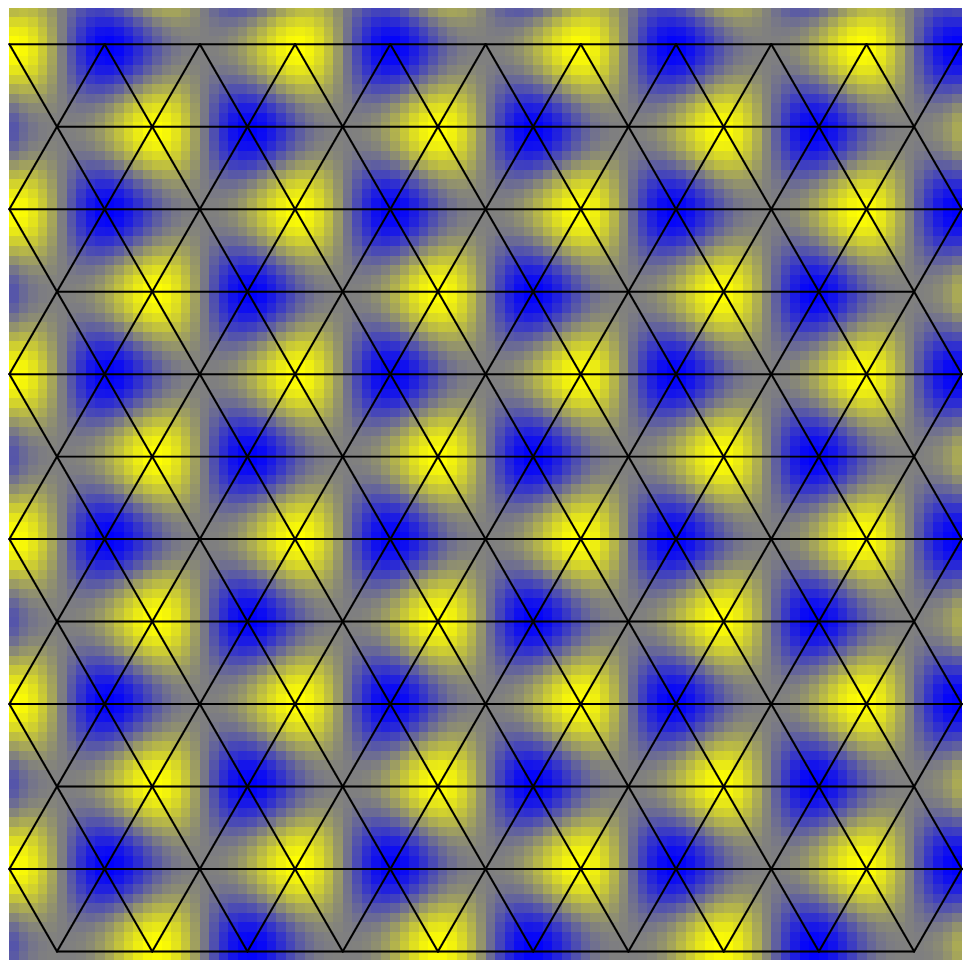}
\includegraphics[width=4cm]{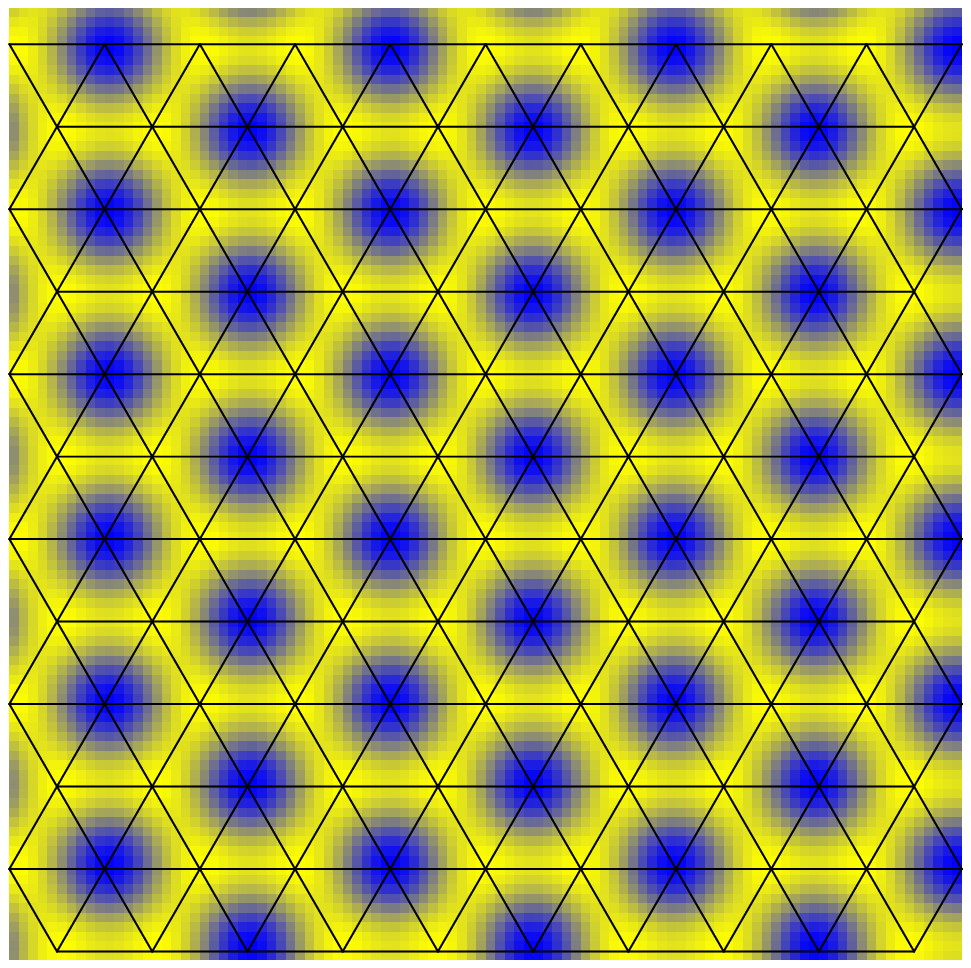}
\caption{Possible charge ordering patterns in the SS3 phases.
Left: ``antiferromagnetic'' supersolid ($w_3 > 0$). 
Right: ``ferrimagnetic'' supersolid ($w_3 < 0$).} 
\label{fig:supersolid}
\end{figure}

Now suppose $t_\theta \gg 1$ but $t_z \ll 1$.  In this limit, we
expect the XY variables condense.  This is a ``Higgs''
phase\cite{Kogut} from the point of view of the $Z_2$ gauge variables:
the linear coupling to $\cos(\Delta_\mu \theta_i/2)$ means that the
$\sigma_{i\mu}$ fields can be regarded as having some non-vanishing
expectation value in this state.  The CP$^1$ fields however remain
uncondensed, and $A_{i\mu}$ remains gapless, so this state retains
superfluidity.  It does, however, break spatial symmetry.  To
understand the nature of this symmetry breaking, let us take for
simplicity $t_z=0$, and again imagine
``summing out'' the $Z_2$ gauge fields while varying $t_\theta$.  The
effective Lagrange density is then
\begin{eqnarray}
  \label{eq:LeffXY}
  {\cal L}_{XY} & = &  V(\Delta_\mu \theta_i) 
  +\frac{1}{2e^2}(\epsilon_{\mu \nu \lambda}
  \Delta_{\nu} A_{i \lambda})^2
\end{eqnarray}
with $V(\Theta) = - \ln \cosh \left[ \frac{t_\theta}{\sqrt{2}}
  \sqrt{1+ \cos(\Theta)}\right]$.  This somewhat unconventional
gradient term has all the same symmetries as the usual
$\cos(\Delta_\mu \theta_i)$ term in an XY model, and indeed reduces to
that form for small $t_\theta$.  On increasing $t_\theta$, therefore,
a 3D=2+1 dimensional XY transition is expected, into a state with a
non-zero expectation value of $e^{i\theta}$.  Comparison with
Eq.~(\ref{eq:psidef}) indicates that $\psi\sim e^{i\pi/4}
e^{-i\theta}$ is an order parameter for this transition.  Note that
$e^{i\theta_i/2}$ is {\sl not} an order parameter since it is not
($Z_2$) gauge invariant.  As noted earlier, $\psi$ is exactly the
order parameter identified in Refs. \onlinecite{Melko05,Damle05,Troyer05}
as characterizing the supersolid phase of the XXZ model on the triangular
lattice.  This supersolid has a three-sublattice structure, so we will
denote it by SS3.  Actually there are two different ordering patterns
possible within the tripled unit cell, depending upon the sign of the
$6$-fold anisotropy term, $w_3$, which should be added to
Eq.~(\ref{eq:LeffXY}).  They are shown in Fig.~\ref{fig:supersolid}.

\begin{figure}[htbp]
  \centering
  \includegraphics[width=4cm]{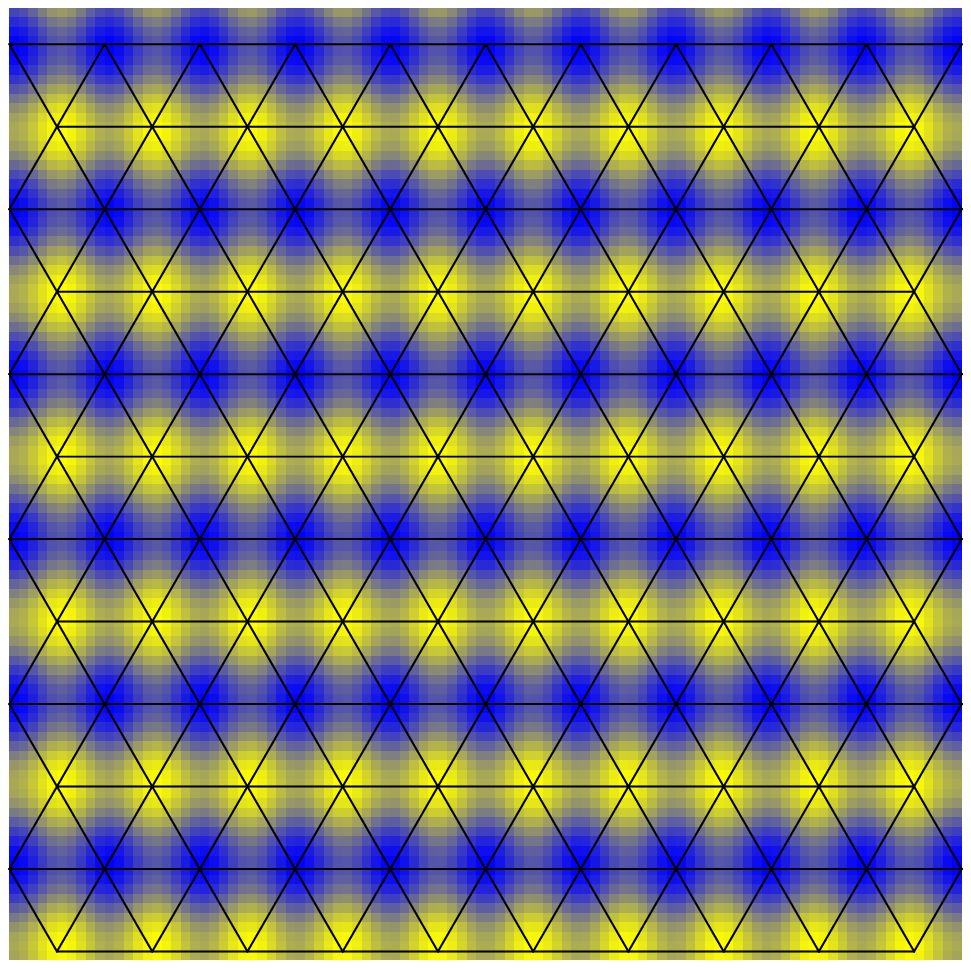}
  \includegraphics[width=4cm]{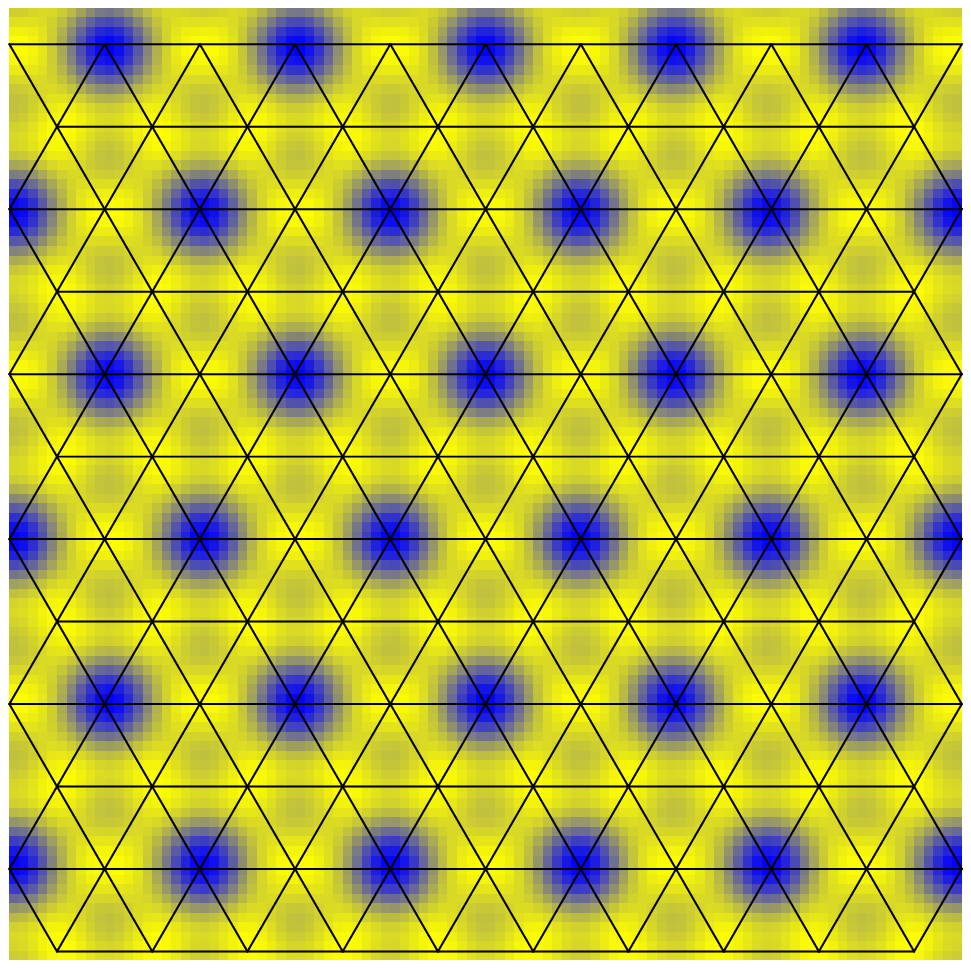}
  \caption{Possible charge orderings in the SS2 phases. Left: ${\bf S}
  = (100)$.  Right: ${\bf S}=(111)$.}
  \label{fig:supersolid2}
\end{figure}

Finally, consider similarly the situation when $t_\theta \ll 1$ but
$t_z$ varies from small to large.  For large $t_z$, we then expect the
CP$^1$ variables order but the XY variables remain uncondensed.
Analogously to the previous case, imagine increasing $t_z$ from small
to large with $t_\theta=0$.  One can again integrate out the Ising
gauge field to obtain an effective action which is $Z_2$-gauge
invariant.  In this case, there are two distinct types of ``kinetic''
terms which arise on nearest-neighbor bonds.   For small $t_z$, they
take the form
\begin{eqnarray}
  \label{eq:LeffSU2}
  {\cal L}_{SU(2)} & = &  - t_S {\bf S}_i \cdot {\bf S}_j - t_2
  e^{-2iA_{i\mu}} z_{i\sigma}^* z_{i\sigma'}^*
    z_{i+\mu,\sigma}^{\vphantom*}z_{i+\mu,\sigma'}^{\vphantom*} 
 \nonumber \\  && +\frac{1}{2e^2}(\epsilon_{\mu \nu \lambda}
  \Delta_{\nu} A_{i \lambda})^2,
\end{eqnarray}
where $t_S,t_2 \sim t_z^2$ and ${\bf S}_i = z_{i}^*
{\boldsymbol\tau}z_i^{\vphantom*}$.  Two distinct types of
``orderings'' are clearly possible on increasing $t_z$.  The most
natural possibility, driven by $t_S$, is for ${\bf S}$ to order.  As
above, this is a supersolid, but with a different set of possible
charge order patterns, characterized by the zone boundary center
wavevectors rather than those at zone corners.  An alternative
possibility, driven by $t_2$, is that the vortex pair field
$z_{i\sigma} z_{i\sigma'}$ condenses.  Such a paired vortex condensate
is a spin-liquid insulator (see Ref. \onlinecite{Balents00}) 
(a ``$Z_2$'' spin liquid in the now-conventional 
nomenclature\cite{SenthilFisher,Wen91}).
The $A_{i\mu}$ gauge fluctuations will tend to suppress such pair
field condensation, so we expect the supersolid phase with $\langle
{\bf S}\rangle\neq 0$ to occur first on increasing $t_z$.  Such
supersolid states have a maximum period of $2$ lattice sites along the
principle axes of the triangular lattice (as can be seen from the
behavior of translations in Eqs.~(\ref{eq:psv}), so we denote these
phases by SS2 (they may have doubled or quadrupled unit cells,
depending upon the orientation of the pseudospin vector).  The two
different ordering patterns for different signs of $w_2$ are shown in
Fig.~\ref{fig:supersolid2}.  

\begin{figure}[htbp]
  \centering
    \includegraphics[width=4cm]{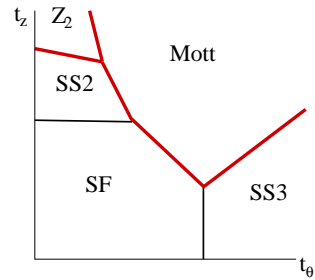}
  \caption{Schematic phase diagram of the $Z_2$ hard spin model.
Thin lines denote conventional LGW-type transitions, thick (red online)
lines denote non-LGW quantum critical points.}
  \label{fig:hspd}
\end{figure}

Putting the different limits of this analysis together and making the
simplest possible interpolation, we expect the phase diagram in
Fig.~\ref{fig:hspd}.  The Mott state may be reached from the superfluid in at least three
distinct ways: by a direct transition described by the continuum
vortex Lagrangian in the previous section, or via two distinct
intermediate supersolid phases.  The transitions from the superfluid
to the two supersolids are ``conventional'', i.e. of LGW type, since
they are described by ordering of the gauge-invariant order parameters
$\psi$ and ${\bf S}$.  The transitions from the supersolids, by
contrast, are {\sl unconventional}.  This is clear from the fact that
the Mott insulator differs from either supersolid by breaking
{\sl more} spatial symmetry {\sl and} by having no off-diagonal
long range order, i.e. by being non-superfluid.  Thus two
symmetry-unrelated order parameters must change in these transitions.
We will return to the nature of these transitions after first
discussing the elementary excitations of the different phases.

\subsection{Elementary excitations}
\label{sec:elem-excit-parall}

\subsubsection{Superfluid phase}
\label{sec:superfluid-phase}

The hard spin model is convenient for describing the elementary
excitations of the phases discussed above.  First consider the
superfluid.  In this case, the elementary excitations are simply
vortices, and the vortex field theory of the previous two sections
already gives a description of the elementary vortex multiplet,
consisting in this case of $4$ vortex flavors (carrying
pseudospin-$1/2$ and XY ``charge'' $\pm 1/2$).  This
should be reproduced by the hard spin model.  Na\"ively, the
``particles'' of the hard spin model are created separately by the
$z_\sigma$ and $e^{\pm i \theta/2}$ fields, so carry only one or the
other of pseudospin or XY charge.  However, in the superfluid region
of Fig.~\ref{fig:hspd}, the $Z_2$ gauge charges (whose interactions
are mediated by $\sigma_{i\mu}$) are {\sl confined}, so the true
elementary excitations are $Z_2$ gauge-neutral bound states $\sim e^{\pm i
  \theta/2} z_\sigma$ which have precisely the appropriate quantum
numbers of the $z_{\pm\sigma}$ vortices.

\subsubsection{Mott phase}
\label{sec:mott-phase}

Next consider the Mott state.  In reality this comprises a number of
different phases, depending upon the signs of the anisotropies
$w_2,w_3$.  However, near to the superfluid-Mott transition, the
latter terms are small, and these distinct phases are approximately
unified into one continuous manifold.  It is useful to discuss the
elementary excitations therefore in the same approximation, and
afterward describe how they are modified once anisotropy is included.
From the point of view of the $Z_2$ hard spin model,
Eq.~(\ref{eq:Lz2}), the Mott state is a Higgs phase, with both XY and
CP$^1$ fields condensed.  The $Z_2$ gauge field can be regarded, in a
choice of gauge, as uniform $\sigma_{i\mu}\approx 1$, in the ground
state.  If we neglect the possibility of deforming the $Z_2$ gauge
field in excited states, there are then two ``obvious'' topological
excitations -- time independent solitons that behave as quantum
particles -- corresponding to textures in the CP$^1$ and XY fields.

First consider the CP$^1$ field.  At spatial infinity, $z_{i\sigma}$
must vary slowly in space to maintain minimal action, as must
$A_{i\mu}$ for the same reason.  We may therefore take a continuum
limit of the first term in Eq.~(\ref{eq:Lz2}), and, taking
$\sigma_{i\mu}\rightarrow 1$, one finds the Lagrangian
\begin{eqnarray}
  \label{eq:Lcp1cond}
 {\cal L}_z & \approx & \frac{1}{2} \varrho_s |\partial_{\mu} {\bf
   S}|^2 + \kappa ({\cal C}_{\mu} - A_{\mu})^2 \nonumber \\
 & & + \frac{1}{2e^2}(\epsilon_{\mu \nu \lambda}
  \partial_{\nu} A_{i \lambda})^2
\end{eqnarray}
with $\varrho_s,\kappa \sim t_z$, and
\begin{equation}
\label{eq:58}
{\cal C}_{\mu} = - \frac{i}{2} 
\left(z^*_{\sigma} \partial_{\mu} z^{\vphantom*}_{\sigma} - 
\partial_{\mu} z^*_{\sigma} z^{\vphantom*}_{\sigma}\right).
\end{equation}
Minimum action configurations therefore have $\partial_\mu {\bf S}=0$
and $A_\mu={\cal C}_\mu$ at infinity.  This requires that, at
infinity, only the phase of the spinor varies, i.e. $z_\sigma =
\xi_\sigma e^{i\Theta}$, where $\xi_\sigma$ is a constant normalized
spinor, and $\Theta$ may depend upon the polar angle from the origin.
At infinity, then
\begin{equation}
  \label{eq:Cinf}
  {\cal C}_\mu = \partial_\mu \Theta.
\end{equation}
Single-valuedness of $z_\sigma$ allows topologically non-trivial
configurations in which $\Theta$ winds by an integer multiple of
$2\pi$, whence
\begin{equation}
  \label{eq:fluxq}
  \oint dx_\mu {\cal C}_\mu = \oint dx_\mu A_\mu = 2\pi n_s,
\end{equation}
the integer $n_s$ being a topological index.  Thus the dual flux of
such configurations is quantized in units of the dual flux quantum.
Since the dual flux measures physical charge, such solitons are
particle-like excitations of the Mott state with physical integral
boson charge $n_s$.  While at infinity the  spinor varies only through
$\Theta$, (and hence the pseudospin ${\bf S}$ is constant),
this cannot hold everywhere in space, since such a ``vortex'' in
$\Theta$ must have a singularity somewhere.  If the spinor is assumed
to vary everywhere slowly in space, so that a uniform continuum limit
can be taken everywhere, then the singularity is avoided by having the
associated amplitude vanish at the skyrmion's ``core'', e.g. in a
configuration of the form
\begin{equation}
  \label{eq:skyex}
  z_\sigma = f(r) e^{i\Theta} \xi_\sigma + \sqrt{1-f(r)} \eta_\sigma,
\end{equation}
where $\xi_\sigma$ and $\eta_\sigma$ are two normalized orthogonal
constant spinors, $\xi_\sigma^* \eta_\sigma^{\vphantom*}=0$, and $f(r)
\rightarrow 1$ as $r\rightarrow \infty$, $f(r) \rightarrow 0$ as
$r\rightarrow 0$ ($r$ is the radial coordinate from the skyrmion
center).  The non-collinear variation of $z_\sigma$ indicates a
non-trivial texture of the pseudospin in the skyrmion.  Quite generally,
if $z_\sigma$ is analytic, one can show that
\begin{equation}
  \label{poynt} n_s = \frac{1}{4\pi} \int d^2 r \, {\bf S} \cdot
  \partial_x {\bf S} \times \partial_y {\bf S},
\end{equation}
directly relating the skyrmion number to the pseudospin texture.
Because the pseudospin itself is constant at infinity, it is apparent
that the skyrmion has a finite size, in the example of
Eq.~(\ref{eq:skyex}) determined by the range of significant spatial
variation of $f(r)$.  This scale is dependent upon details of the
Hamiltonian within the Mott phase, and in general the above
description of the spatially-varying pseudospin is valid only if this
scale is much larger than the unit cell of the Mott charge ordering
pattern.  Deep in the Mott phase (i.e far from the Mott transition),
this may not be the case, and in that case there is not necessarily
any sharp meaning to the pseudospin texture.  In this sense, the
skyrmion/antiskyrmion can be considered as adiabatically connected to
a simple, and patently non-topological, vacancy or interstitial defect
of the Mott ``solid''.  

Near to the Mott transition, however, the
skyrmion is expected to be large, as we now show.  The size of the
skyrmions is determined by the balance of the anisotropy energy ${\cal
  L}_2$ and the interaction energy $(\epsilon_{\mu \nu \lambda}
\partial_{\nu} A_{\lambda})^2$.  A rough estimate for the skyrmion
size can be obtained by a simple dimensional analysis.  The anisotropy
energy of a skyrmion of size $\lambda$ is of the order of $w_2 |S|^4
\lambda^2$, where $|S|$ is the unrescaled amplitude of the pseudospin
vector order parameter.  On the other hand, the interaction energy is
of the order $1/e^2 \lambda^2$.  The optimal skyrmion size is
therefore given by:
\begin{equation}
  \label{eq:63}
  \lambda \sim (w_2 |S|^4 e^2)^{-1/4}.
\end{equation}
Close enough to the critical point, since $|S|$ becomes small, the
skyrmion becomes large, and thus develops a topological character.

Now consider the excitations of the XY field.  Taking $\sigma_{i\mu}
\approx 1$ in Eq.~(\ref{eq:Lz2}), the na\"ive topological excitation
consists of winding $\theta$ at infinity by an integer multiple of
$4\pi$ -- not $2\pi$, since the $\cos(\Delta_\mu \theta_i/2)$ is
not $2\pi$-periodic.  These excitations are {\sl paired} vortices in
the 3-sublattice supersolid order parameter $\psi$.  They are neutral,
since there is no coupling to the dual gauge field.  Like an ordinary
neutral superfluid vortex, they cost a logarithmic energy, neglecting
the XY anisotropy term $w_3$.  When it is included, such a
(double-strength) $\psi$
vortex becomes linearly confined, and converts at long distances to an
intersection point of 12 (!) domain walls, the phase winding
coalescing into these 12 walls radiating outward from the ``vortex''
core.  

Allowing for a non-trivial texture in the $Z_2$ gauge field, however,
a third kind of excitation is possible in the Mott phase.  In
particular, one may consider a ``vison'' or $Z_2$ vortex, around a
point around which any line product of Ising gauge fields gives $-1$.
This requires the existence of a ``cut'', a ray emanating outward from
the vison along which Ising gauge fields crossing the ray are taken
negative (nevertheless, the $Z_2$ flux is non-trivial only through one
plaquette at the vison core).  Such a cut effectively introduces
anti-periodic boundary conditions for both $z_\sigma$ and
$e^{i\theta/2}$ across the cut.  That such a configuration is possible
is of course evident from the definition of $z_\sigma$ and $\theta$ in
Eq.~(\ref{eq:43}), since the apparent discontinuities in $z_\sigma$
and $e^{\pm i \theta/2}$ do not affect the elementary $z_{\pm\sigma}$
fields.  The anti-periodic boundary condition forces topological
defects into both the XY and CP$^1$ fields.  In the XY sector, it
requires the existence of a $\pm 2\pi$ vortex in $\psi$.  As for the
$\pm 4\pi$ vortex above, this costs logarithmic energy neglecting
$w_3$, and degenerates into a linearly confined ``source'' for (in
this case 6) radial domain walls.  In the CP$^1$ sector, it is
slightly less intuitive.  One might have expected the occurrence of
some sort of ``half-skyrmion'' pseudospin texture.  However, this is
not the case.  Anti-periodic boundary conditions require the
discontinuity to persist all the way down from infinity to the vison
core.  There is thus no way to ``relax'' the winding singularity at
infinity into a smooth pseudospin texture.  Instead, the minimal
energy configuration has ${\bf S}$ spatially constant everywhere, i.e.
has the form $z_\sigma=\xi_\sigma e^{i\Theta}$, where $\Theta$ winds
by $\pm \pi$ and $\xi_\sigma$ is constant everywhere save in some
small core region of microscopic size.  However, the continuum action
Eq.~(\ref{eq:Lcp1cond}) still obtains at infinity, so that finite
energy configurations still satisfy Eq.~(\ref{eq:Cinf}) and
$A_\mu={\cal C}_\mu$ at infinity.  Hence, these CP$^1$ ``half-vortex''
configurations carry {\sl fractional boson charges} $\pm 1/2$.  So
by taking into account $Z_2$ gauge vortices, we find a third class of
``elementary'' excitations in the Mott insulator, ``half bosons'' with
a texture in the 3-sublattice supersolid order parameter $\psi$.
These are linearly confined beyond some length at which the 6-fold XY
anisotropy becomes significant.

Of the three types of topological excitations discussed, it is
interesting to note that only the charge $\pm 1$ skyrmion remains
unconfined at the longest scales.  

\subsubsection{SS3 phase}
\label{sec:ss3-phase}

Let us now turn to the SS3 phase, which is described by
Eq.~(\ref{eq:Lz2}) at large $t_\theta$.  As discussed above, the
ground state in this limit can be regarded as a $Z_2$ Higgs phase,
with the $z_\sigma$ field uncondensed.  Hence there are two types of
topological defects: ``paired'' XY vortices in $\psi$, in which
$\theta$ winds by a multiple of $4\pi$ and single XY vortices, in
which $\psi$ winds by $\pm 2\pi$, accompanied by a vison.  Both cost
logarithmic energy at short scales, crossing over to linear
confinement as do similar excitations in the Mott state.  Finally,
there are the CP$^1$ ``particles'' created by the $z_\sigma$ field,
which can be considered to propagate coherently since the $Z_2$ gauge
field is in a Higgs phase.  The single XY vortices have a statistical
interaction with the CP$^1$ quanta, but this does not lead to
significant effects upon the CP$^1$ particles since the XY vortices
are anyway linearly confined.  The CP$^1$ quanta still carry unit dual
gauge charge, and so should be regarded as the {\sl physical}
superfluid vortex excitations of the supersolid.  

\subsubsection{SS2 Phase}
\label{sec:ss2-phase}

The pseudospin vector order parameter ${\bf S}=z^* {\boldsymbol \tau} 
z^{\vphantom*}$ is condensed in the SS2 phase.  However, it is {\sl
  not} a Higgs phase for the $z_\sigma$ fields, since, for instance,
it is still superfluid, i.e. the dual gauge field remains gapless.
Thus in the SS2 phase $Z_2$ quanta are strongly confined, and the
elementary excitations must be $Z_2$ singlets.  One class of
excitations are skyrmions in ${\bf S}$.  They do not carry any
well-defined charge since the boson number conservation symmetry is
anyway broken in the supersolid.  The other quanta are physical
vortices, which are bound states of $z_\sigma$ and $e^{\pm i\theta/2}$
particles, essentially the original $z_{\pm\sigma}$ vortices of the
superfluid.  However, because of the broken pseudospin symmetry in
this phase, there is a preferred pseudospin polarization, and the two
$\sigma$ components of the vortex spinor (choosing a quantization axis
along ${\bf S}$) are no longer energetically equivalent.  Therefore
there is only a two-fold rather than four-fold low-energy vortex
multiplet $\xi_\pm \sim z_{\pm 0}$, taking $\sigma=0$ as the
lower-energy spinor.  

\subsection{Supersolid-Mott quantum critical points}
\label{sec:supers-mott-quant}

\subsubsection{SS3-Mott transition}
\label{sec:ss3-mott-transition}

The supersolid to Mott insulator QCPs are manifestly not of LGW type,
if they occur at all as continuous transitions.  Nevertheless, at
least some aspects of these QCPs can be straightforwardly analyzed by
application of Eq.~(\ref{eq:Lz2}).  Consider first the SS3-Mott
transition.  It is useful to approach the transition first from the
SS3 phase.  Since it can be regarded as a $Z_2$ Higgs state, this is
particularly simple.  In particular, we can treat
$\sigma_{i\mu}\approx 1$ as a constant at low energies.  The $e^{\pm
  i\theta_i/2}$ fields can be regarded as condensed, and moreover at
low energies there are no associated gapless Goldstone modes due to
the 6-fold anisotropy term $w_3$ in Eq.~(\ref{eq:37}). Hence at low
energies, the only important fluctuations are those of the CP$^1$
spinor and the dual gauge field.  The natural critical theory, kept
lattice regularized for simplicity, and neglecting for the moment
pseudospin anisotropy, is thus just
\begin{eqnarray}
  \label{eq:SS3Mott}
  {\cal L}_{NCCP1} & = & - t \sigma_{i\mu} e^{iA_{i\mu}} z^*_{i 
    \sigma} z_{i+\mu \sigma}  +  \frac{1}{2e^2}(\epsilon_{\mu \nu \lambda} \Delta_{\nu} A_{i
    \lambda})^2.   \nonumber \\   
\end{eqnarray}
This is the Non-Compact CP$^1$ (NCCP$^1$) theory studied numerically
in Ref.~\onlinecite{Lesik}, and believed to represent a distinct
universality class of critical phenomena.  It also has a more
intuitive interpretation: it describes the behavior of the $2+1=3$
dimensional $O(3)$ transition associated with ${\bf S}$ {\sl when
  ``hedgehog'' defects in ${\bf S}$ are completely suppressed in the
  partition function}.  Let us now see how this is understood
approaching the QCP from the Mott side.  In this phase, the pseudospin
vector is ordered, but fluctuates more and more strongly as the SS3
phase is approached.  One would expect skyrmion defects to become more
and more prevalent as fluctuations in the pseudospin increase. As
described above in Sec.~\ref{sec:mott-phase}, however, {\sl skyrmions in
the Mott state carry physical boson charge}, so boson number
conservation requires that skyrmion-number ($n_s$ in
Eq.~(\ref{poynt})) must also be conserved.  Happily, skyrmion
number-changing events are exactly the hedgehog defects of the $O(3)$
model, so we see that charge conservation causes this transition to be
of the NCCP$^1$ type.  

Eq.~(\ref{eq:SS3Mott}) and the subsequent discussion neglect
pseudospin anisotropy.  While it is quite likely such anisotropy terms
are irrelevant the NCCP$^1$ fixed point, this requires further study.
Using the hard-spin PSG transformations, Eqs.~(\ref{eq:hstfs}), the
leading anisotropy terms can be shown to be
\begin{eqnarray}
  \label{eq:nccpanis}
  {\cal L}'_{NCCP1} & = & w_2 \!\!\!\! \sum_{a=x,y,z}\!\! (S^a)^4 + w'_2 \left( {\rm Im}\,
  \psi^3 \right) S^x S^y S^z.
\end{eqnarray}
The latter term is allowed by symmetry, but vanishes for $w_3>0$, in
which case $\psi^3$ is purely real.  Note that the perturbations $w_2,w'_2$ in
Eq.~(\ref{eq:nccpanis}) are $8^{\rm th}$ and $6^{\rm th}$ order in the
CP$^1$ fields, respectively, so it is quite plausible that both are
irrelevant at the NCCP$^1$ point, though clearly this is most likely
for $w_3>0$, when the $w'_2$ term is absent.  

A further complication in the case $w_3<0$ is that the non-vanishing
$\psi$ order parameter in this case breaks particle-hole symmetry $C$
(actually the supersolid with $w_3>0$ also breaks C, but preserves the
combination $C \circ T_2 \circ R_{2\pi/3} \circ I_{d_1} \circ I_{d_2}$,
which is sufficient).  Since a supersolid, like a superfluid, is
compressible, this has the difficulty that it implies a non-vanishing
deviation of the spatially averaged density from half-filling (working
at fixed chemical potential chosen to maintain particle-hole symmetry --
i.e. zero Zeeman field in the XXZ model).  In the canonical ensemble,
fixing the average density at $f=1/2$, it implies phase separation.
Formally, this is described in the dual theory by the allowed coupling
term
\begin{equation}
  \label{eq:noph}
  {\cal L}'' = \lambda \left( {\rm Im}\, \psi^3 \right) (\Delta_x A_y -
  \Delta_y A_x),
\end{equation}
since the physical density is the dual magnetic flux.  This indeed leads
to a density deviation from half-filling away from the NCCP$^1$ critical
point, since it leads to a minimum of ${\cal L}_{NCCP1}+{\cal L}''$ with
non-zero $\delta n \sim (\Delta_x A_y - \Delta_y A_x)/2\pi$.  As the
Mott transition is approached, however, the system becomes increasingly
less compressible, and the compressibility certainly vanishes when
superfluidity does.  It is not entirely clear to us how this is resolved
-- the complications are similar to (but more difficult due to the
pseudospin structure) those occuring in the theory of the
normal-superconducting thermal phase transition in a three-dimensional
superconductor in a weak applied external field $H$.\cite{NelsonSeung}
It is possible that, at fixed chemical potential, the NCCP$^1$ critical
fixed point is ``weakly avoided'' at long scales by this effect, most
likely by introducing a narrow region of an ``SS6'' phase -- a
supersolid with the same symmetry as the Mott insulator but with ODLRO
-- between the Mott insulator and ferrimagnetic SS3 state.  Working at fixed
density $f=1/2$, one expects to pass through the NCCP$^1$ point, which
coincides with the critical endpoint of the phase separation region.

\subsubsection{SS2 to Mott transition}
\label{sec:ss2-mott-transition}

As indicated in Sec.~\ref{sec:ss2-phase}, the SS2 phase should be thought
of as a state in which $\langle {\bf S}\rangle\neq 0$, but vortices
themselves are not condensed.  It is not, however, a Higgs phase of
Eq.~(\ref{eq:Lz2}).  Moreover, the important elementary excitations of
this phase are just vortices, which are bound states of the hard-spin
fields.  Therefore it is advantageous to return to the
original soft-spin vortex formulation, and procede by just adding the
term
\begin{equation}
  \label{eq:ss2vect}
  {\cal L}'_{SS2} = -\lambda \langle {\bf S}\rangle \cdot ({\bf S}_+ +
  {\bf S}_- ),
\end{equation}
where of course the ${\bf S}_\pm$ fields on the right should be
understood as composites of $z_{\pm\alpha}$.  As discussed in
Sec.~\ref{sec:ss2-phase}, this splits the 4-fold vortex multiplet into
two 2-fold multiplets.  Taking $\langle {\bf S}\rangle$ along $(100)$
(we will not discuss the $(111)$ case in any detail, but it is similar),
we obtain the {\sl scalar} low energy fields $z_\pm$, defined by
$z_{\pm\alpha} = z_\pm \eta_\alpha$, with $\tau^x \eta=+\eta$ (other
orientations are solved by the obvious generalization).  By considering
the residual symmetry operations of this SS2 state (see
Appendix~\ref{sec:resid-symm-ss2}), we may thereby derive the continuum
action for these two fields:
\begin{eqnarray}
  \label{eq:LSS2Mott}
 &&  {\cal L} =\sum_{\alpha = \pm}
  \left[|(\partial_{\mu} - i A_{\mu}) z_\alpha|^2 + 
    s |z_\alpha|^2\right] + u (z_\alpha^*
    z_\alpha^{\vphantom*})^2 \\ 
    &&- v |z_+|^2 |z_-|^2  + \lambda \,{\rm Im}\, (z_+^* z_-^{\vphantom*})^6+\frac{1}{2 e^2} 
    (\epsilon_{\mu \nu \lambda}\partial_{\nu} A_{\lambda})^2. \nonumber
\end{eqnarray}
Here we have kept the $12^{\rm th}$ order $\lambda$ term because it
is the lowest order term which breaks the staggered $U(1)$ symmetry.  

Remarkably, Eq.~(\ref{eq:LSS2Mott}) is extremely similar to the NCCP$^1$
Lagrangian, differing mainly in that the SU(2) symmetry of that theory
is here reduced by the $v$ term to U(1).  For $v>0$, corresponding to
the easy-plane case, it is the continuum theory for the ``deconfined
quantum critical point'' of Refs.~\onlinecite{dcprefs} that describes
the superfluid to VBS transition on the square lattice, with the
modification that the ``clock'' anisotropy $\lambda$ is here 6-fold
rather than 4-fold.  This theory, neglecting the irrelevant $\lambda$
term, is self-dual at the critical point, and can alternatively be
formulated as a theory of the bose condensation of charge $\pm 1/2$
fractional bosons.  These are just the half-boson excitations discussed
in Sec.~\ref{sec:mott-phase} on the Mott state, which carry a direct
U(1) gauge charge.  

On examining the charge ordering pattern in Fig.~\ref{fig:supersolid2},
an interesting question arises.  The symmetry of the $(100)$ SS2 state
is already consistent with a very simple half-filled Mott insulator,
consisting of stripes of charge on alternate lines of sites (along
principle axes of the triangular lattice).  So it is perfectly
conceivable that in some models, one could have a transition from the
SS2 supersolid to a Mott insulator with the same symmetry as the SS2
phase.  One would expect this to be an XY transition, since only the
superfluid $U(1)$ symmetry is broken across the transition.  Why does
our theory not predict this simpler scenario?  

Firstly, we note that deconfined quantum criticality for the SS2 to Mott
insulator transition studied above is perfectly consistent, since the
Mott insulator in question is {\sl not} the one with the same symmetry
as the SS2 phase.  The question remains why we do not see that
possibility as well.  Our interpretation is that, by starting with the
dual field theory for the $z_{\pm\sigma}$ vortices, we have chosen a
restricted set of vortex modes (this particular multiplet), which
describes the natural instabilities of an {\sl isotropic} triangular
lattice superfluid.  Though we have lowered the symmetry already in the
SS2 phase, we have presumed the low energy excitations in this phase
should be taken same vortex multiplet.  If the spatial symmetry breaking
present in the SS2 phase were taken strong, this might not be a good
assumption, and states origination from other vortex multiplets could
cross the $z_\pm$ states in energy, and lead to instabilities to
different Mott states and also different critical behavior.  We leave an
exploration of this idea for future work.

\subsubsection{SS2 to spin liquid transition}
\label{sec:ss2-spin-liquid}

On passing from the SS2 phase to the $Z_2$ spin liquid in
Fig.~\ref{fig:hspd}, a vortex pair field $z_\sigma z_{\sigma'}$ must
condense.  Because in the SS2 phase, the pseudospin rotational symmetry
is already broken, we expect the pair field composed of two spinors
aligned along the ${\bf S}$ axis to describe the condensate.  This is a
one-component field with dual U(1) gauge charge $2$.  Hence we expect
this transition to be described by a massless charge scalar coupled to a
non-compact $U(1)$ gauge field.  This is just dual to the XY model, so
this can be viewed as an XY transition.  In more physical terms, on
passing from the $Z_2$ phase to the SS2 phase, the half-boson
excitations of the spin liquid condense.  It is clear from this
description that the {\sl symmetry} of the spin liquid is the same as
that of the SS2 phase.

\subsubsection{Spin liquid to Mott transition}
\label{sec:spin-liquid-mott}

In the $Z_2$ spin liquid, both ${\bf S}$ and the vortex pair field are
condensed.  It has, however, the symmetry of the SS2 phase.  It can be
viewed as the Higgs phase of the $\sigma_{i\mu}$ gauge field.  To pass
to the Mott state, which does not have topological order, vison
excitations must condense.  The $e^{\pm i \theta/2}$ particles play the
role of the vison.  This can be seen, e.g. from the fact that these
particles have a statistical interaction with the half-bosons in the
$Z_2$ phase, which are $\pi$-flux tubes in $A_{i\mu}$.  The non-standard
feature is that the visons also carry space group quantum numbers --
since the $e^{\pm i \theta/2}$ operator is a ``square root'' of the SS3
order parameter (the $\theta$ transformations are given in
Appendix~\ref{sec:hard-spin-tranf}).  Actually this transition can be
understood from Eq.~(\ref{eq:Lz2}) simply by ``freezing'' $\sigma_{i\mu}
\approx 1$ and treating the $z_\sigma$ fields as condensed.  It is thus
clearlyan XY transition, with either $6$-fold or $12$-fold ``clock''
anisotropy (it is doubled since ``half'' the SS3 order parameter is
condensing), for ${\bf S}$ along $(100)$ and $(111)$, respectively.

\subsection{Anti-parallel pseudospins}
\label{sec:anti-parall-pseud}

Here we {\sl briefly} sketch the results of an analogous study of the
antiparallel pseudospin case.  Using the parameterization in 
Eq.~(\ref{eq:44}) to
define the hard-spin degrees of freedom, and gauging the $Z_2$
redundancy, the appropriate hard spin model is
\begin{eqnarray}
\label{eq:Lz2ap}
  {\cal L}_{\rm Z_2}^{-\parallel} & = & - t_z \sigma_{i\mu} z^*_{i 
    \sigma} z_{i+\mu \sigma} - t_\theta \sigma_{i\mu} \cos(\Delta_\mu
  \theta_i/2 - A_{i\mu}) \nonumber \\
  && +  \frac{1}{2e^2}(\epsilon_{\mu \nu \lambda} \Delta_{\nu} A_{i
    \lambda})^2.
\end{eqnarray}
Note that, in this case, the dual gauge field is coupled to the XY  and
not the CP$^1$ degree of freedom.  

\begin{figure}[htbp]
  \centering
      \includegraphics[width=4cm]{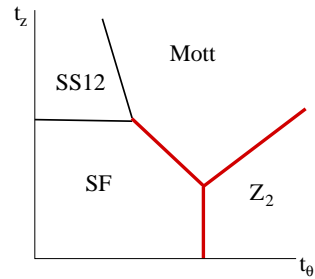}
  \caption{Schematic phase diagram for antiparallel pseudospin vectors.
Thick and thin lines have the same meaning as in Fig.\ref{fig:hspd}.}
  \label{fig:appd}
\end{figure}

Study of various limits and interpolation determines the phase structure
of Eq.~(\ref{eq:Lz2ap}), which is schematically shown in
Fig.~\ref{fig:appd}.  In addition to the superfluid and Mott states
occuring when both $t_z,t_\theta$ are small and large, respectively,
there is a supersolid phase, SS12 (it has a minimal unit cell of $12$
sites), and a $Z_2$ spin liquid insulating phase.  Unlike in the
parallel pseudospin model, the SS12 supersolid has {\sl the same
  symmetry} as the nearby Mott insulator.  This is a direct consequence
of the fact that the dual gauge charge is in this case carried by the
$U(1)$ degree of freedom $e^{i\theta}$, and this does not carry any
space group quantum numbers.  For the same reason, the $Z_2$ spin liquid
phase at large $t_\theta$ does not break {\sl any} symmetries.  Note
that of the four phases in Fig.~\ref{fig:appd}, only the $Z_2$ spin
liquid is ``exotic'', i.e. has an underlying topological order and
unconventional excitations not captured by any local mean-field theory
and order parameters.

The excitations are as follows.  In the Mott state there are {\sl neutral}
skyrmions (since the $z_\sigma$ fields carry no dual gauge charge),
confined charge $\pm 1/2$ excitations (visons bound to half-vortices
in $z_\sigma,\theta$), and charge $\pm 1$ excitations that can be
viewed as $\pm 4\pi$ vortices in $\theta$.  Because $\theta$ is
coupled to $A_{i\mu}$, the latter cost finite energy, and are clearly
adiabatically connected to vacancy/interstitials in the Mott state.  The
SS12 phase has the same skyrmion textures, but no well-defined charge
excitations since it is a superfluid.  Instead, it has a single physical
vortex/antivortex excitation created by $e^{\pm i \theta/2}$.  The
``triviality'' of the vortex multiplet is consistent with the broken
symmetry of the supersolid, whose enlarged unit cell contains on average
an integer number ($6$ in the simplest case) bosons.  The $Z_2$ spin
liquid has physical boson charge $\pm 1/2$ excitations ($2\pi$ vortices in
$\theta$ accompanied by a ``vison'' in $\sigma_{i\mu}$) and physical
``vison'' excitations (created by $e^{\pm i\theta/2}$ and $z_{\sigma}$)
which carry spatial quantum numbers.  

The direct transition from superfluid to antiparallel Mott state is
described by the continuum action of the previous section.  The
superfluid-SS12 and the SS12-Mott critical points {\sl are also
  conventional}, since each is characterized by a change in a single
order parameter.  The superfluid-SS12 transition is described by an LGW
theory for the ${\bf d}$-vector, while the SS12-Mott transition is
simply an XY transition for the superfluid order parameter.  The
superfluid-$Z_2$ transition is also an XY transition, which can be
understood as a condensation of charge $\pm 1/2$ ``half-bosons'' (in
principle this changes universal amplitudes from the conventional
superfluid to integer-filling Mott transition, which is also
XY-like\cite{Sedgewick}).  The $Z_2$ to Mott transition is modeled by
the CP$^1$ action {\sl with no gauge field}, which has the physical
interpretation of modeling vison condensation.  We have not attempted to
consider the effects of various anisotropies on these transitions.

\section{Discussion}
\label{fin}         
 
In this paper we have presented a phenomenological dual vortex theory
of the interplay between Mott localization and geometrical frustration
for interacting bosons at half-filling on the triangular lattice.
This approach reveals a variety of novel quantum phases and phase
transitions which may occur if the superfluid and Mott insulating
states occur in close proximity to one another in phase space.  Of
particular interest are the continuous superfluid-Mott insulator
transition predicted by mean field theory, the two supersolid phases,
and the occurrence of the recently-discovered NCCP$^1$ critical
universitality class at the 3-sublattice supersolid to Mott insulator
transition.  In this discussion, we will provide a more direct
physical picture of some of these phenomena, and address the prospects of
observing them in simple microscopic boson or spin models.

A useful starting point for the discussion is the recent demonstration
that a supersolid phase indeed occurs in the simplest spin-$1/2$ XXZ
model,
\begin{equation}
  \label{eq:xxz}
  H_{XXZ} = \sum_{\langle ij\rangle} -J_\perp (S_i^x S_j^x + S_i^y
  S_j^y) + J_z S_i^z S_j^z,
\end{equation}
with ferromagnetic XY and antiferromagnetic Ising exchanges
($J_\perp,J_z>0$) (equivalently, hard-core bosons with nearest
neighbor repulsion) on nearest-neighbor links of the triangular
lattice.\cite{Melko05,Damle05,Troyer05} This model was shown to be in
a 3-sublattice SS3-type phase for $J_z\gtrsim 5 J_\perp$, and this
phase persists up to and including $J_z=\infty$.  A number of features
of the numerical results on the supersolid at large $J_z$ are notable.
First, although superfluidity survives, it is {\sl extremely} weak, as
characterized by the superfluid stiffness, which is approximately
$250$ times smaller in the large $J_z$ supersolid than in the pure XY
model ($J_z=0$).  Second, it is exceedingly difficult to distinguish
numerically on even relatively large lattices between the two
different types of SS3 charge ordering patterns.
Ref.~\onlinecite{Melko05}\ was unable to distinguish them numerically
by direct measurement of boson density correlation functions, while
Ref.~\onlinecite{Damle05}\ claimed to do so, but only for large
lattices of $18\times 18$ sites with a very small signal.
Furthermore, a deviation of the density from half-filling is
expected\cite{Melko05}\ in the ``ferrimagnetic'' SS3 phase identified in
Ref.~\onlinecite{Damle05}\ (which corresponds in our theory to the one
with $w_3<0$), but appears to be exceedingly minute if observable at
all computationally.  Apparently there is very little splitting
energetically between the two SS3 states, even at $J_z=\infty$, a
point at which there is no intrinsic small parameter in the
microscopic Hamiltonian -- the effective Hamiltonian is simply the XY
exchange projected into the Hilbert space spanned by the manifold of
classical Ising antiferromagnetic ground states on the triangular
lattice.  

The present theory offers a partial explanation for these puzzling
observations.  We interpret the weakness of superfluidity as evidence
that the system is in close proximity to a Mott insulating state.  If
so, our dual vortex field theory, which is built around a superfluid
to Mott insulating transition, should apply.  The tiny energy
splitting between the two SS3 states is then understandable: the
term which dictates this splitting, $w_3$ is $12^{\rm th}$ order in
the basic $z_{\pm\sigma}$ vortex fields, and clearly strongly
irrelevant at the superfluid-Mott QCP.  Furthermore, the smallness of
any spontaneous density deviation from $1/2$-filling, which as
indicated above is expected for the ``ferrimagnetic'' SS3 state, is
also expected, since the density deviation must vanish as the Mott
state, which has density of exactly $1/2$ and is incompressible, is
approached.  This deviation also vanishes at the transition from this
state to the superfluid (as seen in LGW theory\cite{Melko05}), so it
is likely small throughout the ferrimagnetic phase.

Given these arguments for proximity to the Mott state, it seems likely
that only a small perturbation of the XXZ Hamiltonian may be required
to push it into a Mott phase, and in so doing observe the very
interesting NCCP$^1$ criticality at the SS3-Mott quantum critical
point.  Let us try to develop a more physical picture of this
transition.  We will begin by providing a cartoon understanding of the
SS3 phases.  In the case $w_3<0$, as discussed in
Refs.~\onlinecite{Damle05,Troyer05}, the ferrimagnetic SS3 state can
be understood crudely by first forming a ``solid'' of bosons with a
density of $1/3$, with one boson occupying each of the sites of one of
the three $\sqrt{3}\times\sqrt{3}$ triangular sublattices of the
original lattice.  This leaves, at $f=1/2$, a density of $1/4$ boson
per the remaining sites, which form a honeycomb lattice, with the
``solid'' bosons in the centers of the honeycomb plaquettes.  The SS3
state can be viewed as a superfluid of these remaining bosons
(alternatively, one can make the same construction with holes
replacing bosons, leading to a different but equivalent state -- which
illustrates the spontaneously broken particle-hole symmetry of the
ferrimagnet).  In the opposite case $w_3>0$, there is no spontaneous
density polarization, and the ``sublattice magnetizations'' take the
values $\langle S_i^z \rangle =(m,-m,0)$ on the three inequivalent
sites. We will call this the ``antiferromagnetic'' state because of
the exactly opposite Ising moments on two of the three sublattices.
This state can be understood by a similar cartoon.  In particular,
take a honeycomb sublattice of the triangular lattice, and on this
sublattice form a $1/2$-filled Mott insulator of alternating empty and
occupied sites (an Ising N\'eel state in spin language).  The sites at
the centers of the honeycomb plaquettes form a
$\sqrt{3}\times\sqrt{3}$ triangular sublattice, and we put the
remaining bosons into a half-filled superfluid on this sublattice.
These cartoon pictures would clearly tend to favor the ferrimagnetic
state in the XXZ model, since bosons must hop (presumably virtually)
between second neighbor sites to stabilize the triangular superfluid
in the antiferromagnetic state.  This immediately suggests that, to
study the antiferromagnetic supersolid, one needs only to add a second
neighbor XY exchange to the XXZ model,
\begin{equation}
  \label{eq:H2ndneigh}
  H' = -J'_\perp \sum_{\langle\langle ij\rangle\rangle} S_i^x S_j^x + S_i^y
  S_j^y.
\end{equation}

\begin{figure}[htbp]
  \centering
  \includegraphics[width=4cm]{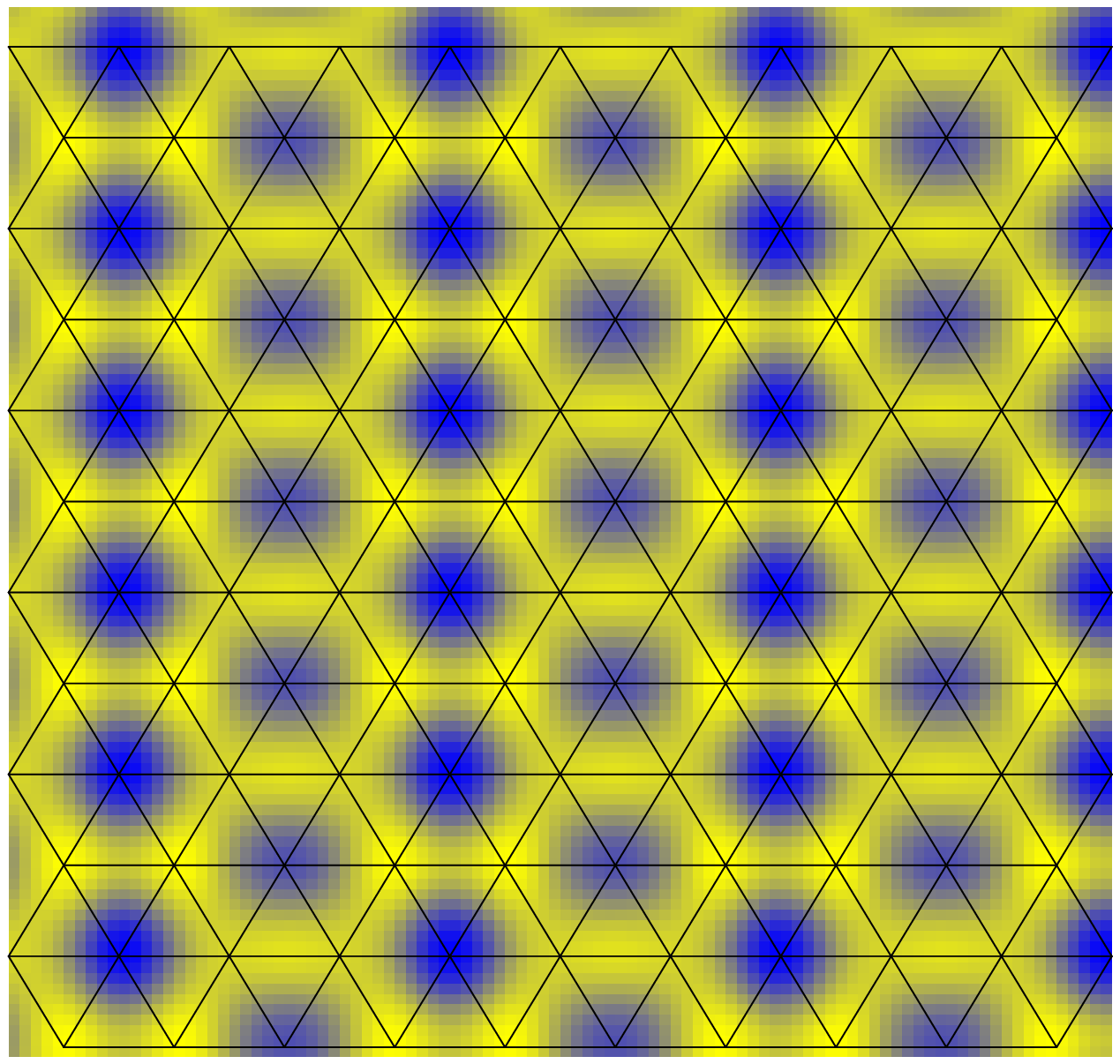}
  \includegraphics[width=4cm]{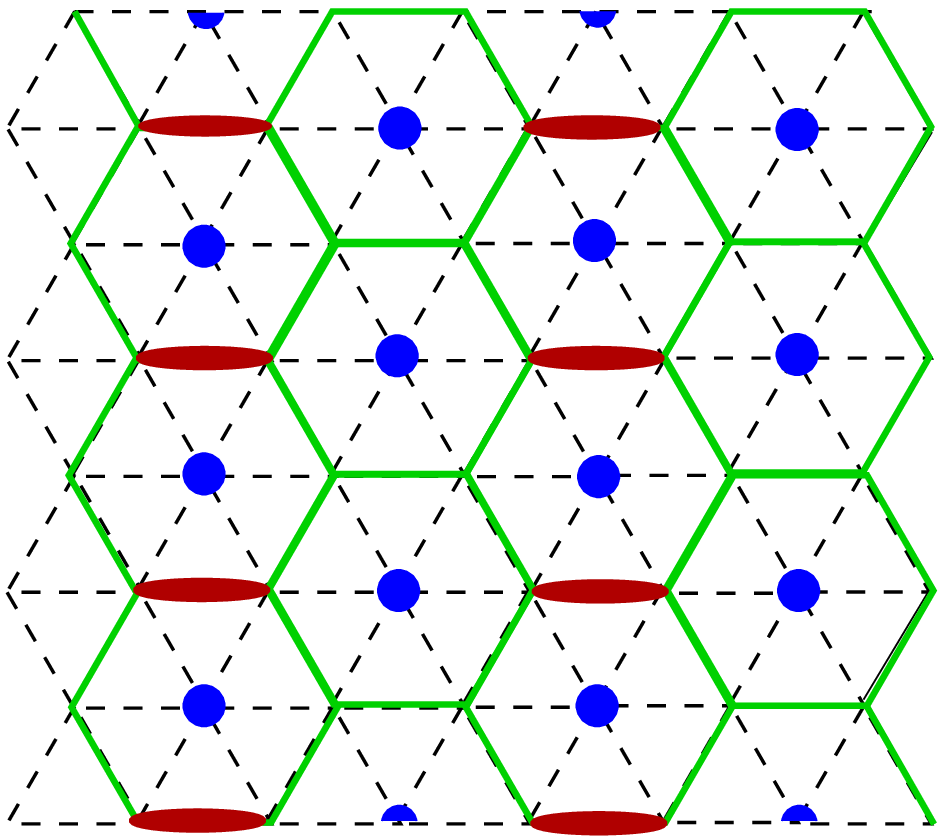}
  \caption{Density plot for the ferrimagnetic solid (left) and the
    corresponding cartoon (right).  In the density plot, the honeycomb
    sublattice is clearly visible.  In the cartoon, the honeycomb
    lattice on which the superfluid lives is overlaid in green, with the
    frozen bosons in blue, and the valence bonds that form on passing
    from the ferrimagnetic SS3 phase to the ferrimagnetic solid are
    drawn as red ovals (color online). }
  \label{fig:MottSS}
\end{figure}

With these pictures of the SS3 phases in hand, it is natural to view the
transition to the Mott insulator as a ``crystallization'' of the
superfluid sublattices of the supersolid.  Indeed, we find that this
provides a simple physical picture consistent with the symmetries of the
appropriate Mott phases.  Consider first the transition from the
ferrimagnet. Here we have a $1/4$-filled honeycomb lattice of bosons,
which is to undergo a superfluid to Mott insulator transition.  At
$1/4$-filling, these bosons are unlikely to form a simple
``crystalline'' Mott insulator, since they would be too widely separated
to substantially interact.  Instead, more likely Mott states are valence
bond solids, in which the bosons resonate between two or more sites
(within still-localized wavefunctions).  The most natural candidate is a
``columnar'' valence bond solid (VBS) state, in which alternating
columns of bonds are occupied by one boson.  Indeed, the state predicted
by the vortex mean field theory for $w_3<0$ and $w_2<0$ has exactly the
symmetries of the columnar valence bond solid, see Fig.~\ref{fig:MottSS}.  
One may also
convince oneself of the validity of the columnar VBS picture by counting
the number of distinct Mott states.  Fix the location of the
$\sqrt{3}\times\sqrt{3}$ superlattice and hence the honeycomb
sublattice.  One can then place the valence bonds along columns parallel
to any of the three principle axes (of the honeycomb, which has
principle axes halfway between those of the underlying triangular
lattice), and for each such orientation, they may lie on even or odd
columns. Hence one expects six states.  This is precisely the number of
distinct choices of vector pseudospin along the $\pm \hat{x},\pm\hat{y},
\pm\hat{z}$ axes.

Next consider the antiferromagnetic supersolid.  We have an effective
$\sqrt{3}\times\sqrt{3}$ triangular sublattice of bosons at
half-filling, living in the plaquette centers of an
``antiferromagnetic'' bose solid on the honeycomb lattice.  In the
absence of this surrounding solid, the triangular sublattice would have,
scaled up to its size, all the same symmetries as the original
triangular lattice.  The formation of a Mott insulator on this
triangular sublattice would thus na\"ively appear to be just as
formidable a problem as the original one.  The staggered solid, however,
breaks $I_{d_2}$ and $C$, preserving only the combination, $I_{d_2}\circ
C$.  This means, were we to repeat the PSG analysis for a dual vortex
theory of the $\sqrt{3}\times\sqrt{3}$ sublattice bosons, there is no
symmetry to prevent the Ising order parameter $\Phi=|z_+|^2-|z_-|^2$
from appearing as a term in the action, breaking the $\alpha=\pm$
``flavor'' degeneracy of the vortex multiplet.  Clearly in such a
theory, then only one of the two flavors will condense, and only those
phases in which one pseudospin is non-zero will appear.  Incidentally,
one may identify therefore the $z_\sigma$ spinor in the hard-spin action
with the lower-energy vortex flavor in this sublattice theory.  The
phases with only a single pseudospin condensed are ``staggered'' phases
of the original bosons.  The simplest of these is just a ``columnar
crystal'', in which bosons on alternating columns of triangular lattice
{\sl sites} (along some principle axis) are occupied and unoccupied.
Superimposing this upon the surrounding antiferromagnetic honeycomb
lattice, one remarkably obtains an ordering pattern again {\sl
  identical} in symmetry to the original mean-field solid with $w_3>0$
and $w_2<0$!  The counting of states is also the same as for the
ferrimagnetic case above, since the ``columns'' have the same set of
orientations and have only shifted from bonds to sites, once again in
agreement with the configurations of ${\bf S}$.  

One can easily extend these constructions to the cases with pseudospin
along (111).  For brevity, we relegate this to
Appendix~\ref{sec:toy-model-wavefunctions}.

\begin{figure}[htbp]
  \centering
  \includegraphics[width=4cm]{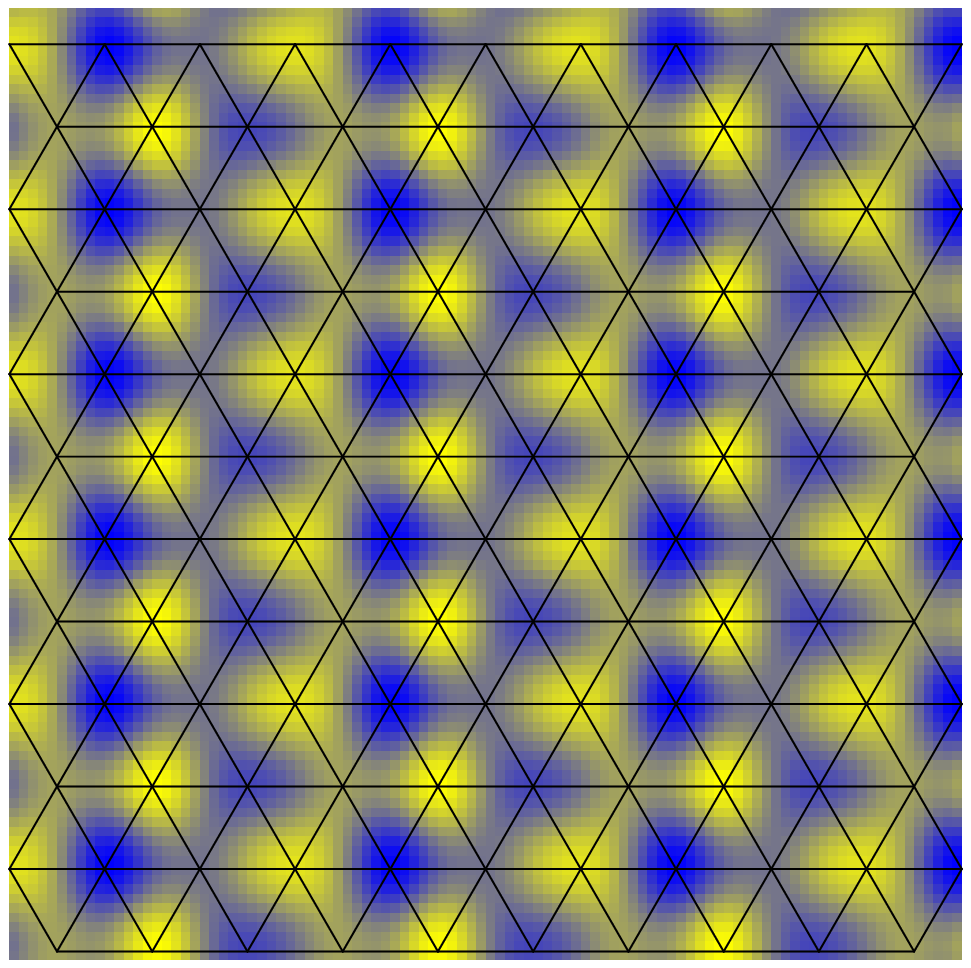}
  \includegraphics[width=4cm]{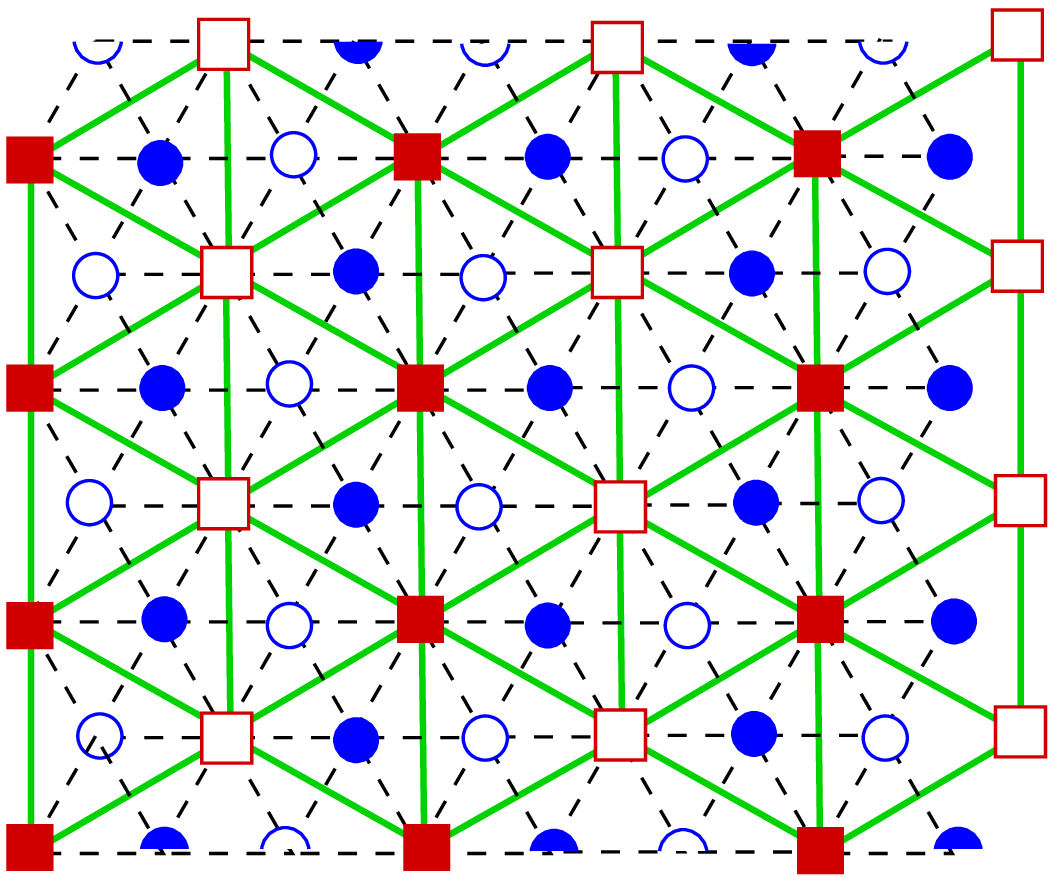}
  \caption{Density plot for the antiferromagnetic solid (left) and the
    corresponding cartoon (right).  In the cartoon, the triangular
    lattice on which the superfluid lives is overlaid in green, with the
    crystallized  bosons/vacancies forming the Mott insulating honeycomb
    lattice in blue (filled and empty circles).  The additional
    bosons/vacancies that solidify  on passing
    from the antiferromagnetic SS3 phase to the antiferromagnetic solid are
    drawn as red squares (filled and empty, respectively). }
  \label{fig:afsolid}
\end{figure}

The cartoons of the ferrimagnetic and antiferromagnetic
supersolids/solids not only reproduce the symmetries of the phases, but
also can be directly used to obtain the critical NCCP$^1$ theory of the
SS3-Mott transition.  In the ferrimagnetic case, the cartoon picture is
a superfluid-Mott transition on a $1/4$-filled honeycomb lattice of
bosons.  In Appendix~\ref{sec:honeycomb-lattice-at}, we sketch how a
dual vortex analysis of that ``cartoon'' problem directly recovers the
appropriate NCCP$^1$ theory for this case.  In the antiferromagnetic
case, a similar argument has already been sketched for the cartoon of an
half-filled triangular $\sqrt{3}\times\sqrt{3}$ sublattice in the
background of the honeycomb antiferromagnetic Mott state, which removes
one of the two pseudospinors from the theory, likewise recovering the
NCCP$^1$ model and the appropriate anisotropy.

It thus seems of considerable interest to pursue small perturbations of
the XXZ model which might take it into a Mott phase, and between
ferrimagnetic and antiferromagnetic states.  To make the latter
transition, from the ferrimagnetic to antiferromagnetic SS3 supersolid,
we can make a simple recommendation: the inclusion of weak
second-neighbor ferromagnetic exchange, as in Eq.~(\ref{eq:H2ndneigh}),
should favor the antiferromagnetic state by allowing bosons to better
move on the triangular sublattice.  The antiferromagnetic states seem
slightly more favorable than the ferrimagnetic ones theoretically as
candidates for observing the NCCP$^1$ criticality, because the latter
preserve particle-hole symmetry in the supersolid, and have weaker
pseudospin anisotropy terms.  Understanding what terms microscopically
would be needed to push the XXZ model toward the insulators is a
nontrivial problem.  However, it seems clear that introducing
further-neighbor Ising exchange will lead to {\sl some} Mott insulating
states, when $J_z$ is large, since they spoil the degeneracy of the
classical Ising manifold.  Exploration of this problem numerically is
tempting.  

Analytically, it is quite interesting how naturally the dual vortex
approach leads to the observed supersolid states.  Coming from the Mott
insulator with parallel pseudospins, the deconfined charge excitations
are skyrmions.  If one imagines lowering the charge gap in the
insulator, it is most natural that these should condense, destroying the
pseudospin order of the insulator and simultaneously initiating
superfluidity.  While other scenarios such as the direct Mott insulator
to superfluid transition are possible theoretically (and described
analytically in this paper), they entail a non-trivial screening
so that confinement of other (fractionally charged) excitations of the
Mott state becomes progressively weaker as the transition
is approached.  Thus the SS3 supersolid is probably the most intuitive
``neighbor'' of the Mott state, and with this understanding its
occurence in the strong interaction limit of the XXZ model is no longer
surprising.  It is interesting to speculate as to whether it might be
possible to formulate the boson-vortex duality used heavily in this
paper directly in this strong-coupling region, i.e. in the classical
Ising antiferromagnetic Hilbert space.  Such a formulation could
potentially provide a quantitative method for attacking problems of Mott
transitions with strong frustration.

\begin{acknowledgments}
We acknowledge useful discussions with O.I.~Motrunich, A.~Paramekanti and 
A.~Vishwanath.
Financial support was provided by the National Science Foundation grant
DMR-9985255 and by the Packard Foundation. 
\end{acknowledgments} 

\appendix
\section{Various PSG details}
\label{sec:psg-momentum-space}

\begin{widetext}
The PSG transformations in momentum space are given by:
\begin{eqnarray}
\label{eq:9}
T_1 & : & \left\{\begin{array}{ccl} \psi_1(k_1,k_2) & \rightarrow &
  \psi_1(k_1,k_2 - 2 \pi f) e^{-i k_1} \\
 \psi_2(k_1,k_2) & \rightarrow & \psi_2(k_1,k_2 - 2 \pi f) e^{-i k_1}
 \omega\end{array}\right. ,\nonumber \\ 
T_2 & : &  \left\{\begin{array}{ccl} \psi_1(k_1,k_2) & \rightarrow &
    \psi_1(k_1,k_2) e^{-i k_2} \\ \psi_2(k_1,k_2) & \rightarrow &
    \psi_2(k_1,k_2) e^{-i k_2} \end{array}\right. ,\nonumber \\
R_{2\pi/3} & : &  \left\{\begin{array}{ccl} \psi_1(k_1,k_2) &
    \rightarrow &  \frac{1}{q} 
\sum_{m,n=0}^{q-1}\omega^{m(m+2n+\nu_q)/2} \psi_1(k_2 + 2 \pi f m, -
k_1 - k_2 + 2 \pi f(n-1/2+\nu_q/2)) \\
 \psi_2(k_1,k_2) & \rightarrow & \frac{1}{q}e^{i(k_1+k_2)}
 \sum_{m,n=0}^{q-1}\omega^{(m-1)(m+2n-1+\nu_q)/2} \psi_2(k_2 + 2 \pi f
 m, - k_1 - k_2 + 2 \pi  f(n-1/2+\nu_q/2)) \end{array}\right. ,\nonumber \\
I_{d_1} & : & \left\{\begin{array}{ccl} \psi_1(k_1,k_2) & \rightarrow
    & \frac{1}{q} \sum_{m,n=0}^{q-1}
\psi_1^*(k_2 + 2 \pi f m, k_1 + 2 \pi f n) \omega^{-mn} \\
\psi_2(k_1,k_2) & \rightarrow & \frac{1}{q} e^{i(k_1+k_2)}
\sum_{m,n=0}^{q-1} \psi_2^*(k_2 + 2 \pi f m, k_1 + 2 \pi f n) 
\omega^{-(m-1)n} \end{array}\right. , \nonumber \\
I_{d_2} & : & \left\{ \begin{array}{ccl}  \psi_1(k_1,k_2) &
    \rightarrow & \frac{1}{q}e^{-i k_1} \sum_{m,n=0}^{q-1}
\psi_2^*(-k_2 + 2 \pi f m, -k_1 + 2 \pi f n) \omega^{-(m-1)n} \\
\psi_2(k_1,k_2) & \rightarrow & \frac{1}{q} e^{i k_2}
\sum_{m,n=0}^{q-1} \psi_1^*(-k_2 + 2 \pi f m, -k_1 + 2 \pi f n) 
\omega^{-mn} \end{array}\right. ,
\end{eqnarray}
where $\nu_q = q - 2[q/2]$.
\end{widetext}

Using these general forms for the PSG in momentum space, 
it is straightforward to obtain
Eqs.~(\ref{eq:PSGtodd},\ref{eq:PSGteven}) of the main text.  The
remaining PSG transformations for the multiplets are:
\begin{eqnarray}
\label{eq:14}
&&R_{2 \pi/3} : \varphi_{\ell} \rightarrow \frac{1}{\sqrt{q}} 
\sum_{\ell'=0}^{q-1}
\omega^{-\ell(\ell+2 \ell'+1)/2} \varphi_{\ell'}, \nonumber \\
&&I_{d_1} : \varphi_{\ell} \rightarrow \frac{1}{\sqrt{q}} 
\sum_{\ell' = 0}^{q-1} \omega^{\ell \ell'} \varphi^*_{\ell'}, \nonumber \\
&&I_{d_2} : \varphi_{\ell} \rightarrow \frac{1}{\sqrt{q}} 
\sum_{\ell' = 0}^{q-1} \omega^{-\ell \ell'} \varphi^*_{\ell'},
\end{eqnarray}
for $q$ odd, and
\begin{eqnarray}
\label{eq:12}
&&R_{2 \pi/3} : \varphi_{\alpha\sigma} \rightarrow e^{-i \pi \alpha \sigma/q} 
\frac{e^{i \eta_1(\alpha,f)}}{\sqrt{q}} \sum_{\sigma'=0}^{q-1}
\omega^{-\sigma(\sigma+2 \sigma'+1)/2} \varphi_{\alpha\sigma'}, \nonumber \\
&&I_{d_1} : \varphi_{\alpha\sigma} \rightarrow \frac{e^{i \eta_2(\alpha,f)}}
{\sqrt{q}} \sum_{\sigma' = 0}^{q-1} \omega^{\sigma \sigma'} 
\varphi^*_{\alpha\sigma'}, \nonumber \\
&&I_{d_2} : \varphi_{\alpha\sigma} \rightarrow \frac{e^{i \eta_3(\alpha,f)}}
{\sqrt{q}} \sum_{\sigma' = 0}^{q-1} \omega^{-\sigma \sigma'} 
\varphi^*_{-\alpha,\sigma'},
\end{eqnarray} 
when $q$ is an even integer.  Appying $R_{2\pi/3}$ to $\tilde
\varrho^{\alpha}_{mn}$, one finds that
\begin{equation}
\eta_1(\alpha) = - \frac{\pi \alpha}{6 q}.
\end{equation}
Applying $I_{d_2}$, one finds that $\eta_3(\alpha) = 0$. 
The remaining phase factor $\eta_2(\alpha,f)$ is
difficult to find analytically in the general case.  In the case
$f=1/2$ considered in detail in the text, it is, however, equal to
$\eta_1(\alpha)$, so collecting these results,
\begin{equation}
\label{eq:13}
\eta_1(\alpha) = \eta_2(\alpha) = - \pi \alpha / 12, \,\,\,
\eta_3(\alpha) = 0\qquad \mbox{for $f=1/2$}.
\end{equation}

It is possible that $\eta_1(\alpha) = \eta_2(\alpha)$ holds at a general 
filling, not just at $f=1/2$, but it is not obvious. 

For the specific case of $f=1/2$ ($p=1,q=2$), we can impose invariance
under the particle-hole transformation,
\begin{equation}
  \label{eq:psgph}
  C\, : \, \varphi_{\pm\sigma} \rightarrow \varphi^*_{\mp\sigma}.
\end{equation}

Imposing invariance under the rotation and reflection operations above
upon the quartic action ${\cal L}_1$ taken in the form of
Eq.~(\ref{eq:16}) yields the set of conditions
\begin{eqnarray}
\label{eq:18}
&&\gamma_{m n} = \gamma_{-m,-n}, \nonumber \\
&&\gamma_{m n} = \gamma_{m,m-n}, \nonumber \\
&&\gamma_{m n} = \gamma_{m-2n,-n}^*, \nonumber \\
&&\gamma_{m' n'} = \frac{1}{q} \sum_{m,n=0}^{q-1} 
\gamma_{m n}  \omega^{m n' - n m'},  \nonumber \\
&&\gamma_{m' n'} = \frac{1}{q} \sum_{m,n=0}^{q-1}
\gamma_{m n} \omega^{n(m'-n-2 n') + m(n+n')}, \nonumber \\
\end{eqnarray}
for $q$ odd, and 
\begin{eqnarray}
\label{eq:17}
&&\gamma_{m n}^{\alpha \beta} = \gamma_{-m,-n}^{-\alpha -\beta}, \nonumber \\
&&\gamma_{m n}^{\alpha \beta} = \gamma_{-m,-n}^{\beta \alpha}, \nonumber \\
&&\gamma_{m n}^{\alpha \beta} = \gamma_{m-2n,-n}^{\alpha \beta *}, \nonumber \\
&&\gamma_{m' n'}^{\alpha \beta} = \frac{1}{q} \sum_{m,n=0}^{q-1} 
\gamma_{m n}^{\alpha \beta} \omega^{m n' - n m'},  \nonumber \\
&&\gamma^{\alpha \beta}_{m' n'} = \frac{1}{q} \sum_{m,n=0}^{q-1}
\gamma_{m n}^{\alpha \beta} e^{i \pi n (\beta - \alpha)/q}
\omega^{n(m'-n-2 n') + m(n+n')}, \nonumber \\
\end{eqnarray}
for $q$ even.

\section{Pseudospin transformations at $f=1/2$}
\label{sec:pseud-transf}

For convenience, we give the transformation properties of the spinor
$z_{\alpha\sigma}$ vortex fields.  With the pseudospin index
suppressed, we find
\begin{eqnarray}
  \label{eq:ztfs}
  T_1&:& \left\{ \begin{array}{lcr}  
      z_{+}  & \rightarrow & e^{-i\pi/6} \tau^x z_+  \\
      z_{-}  & \rightarrow & -e^{i\pi/6} \tau^x z_-
    \end{array}\right. ,\nonumber \\
  T_2&:& \left\{ \begin{array}{lcr}  
      z_{+}  & \rightarrow & e^{-i\pi/6} \tau^z z_+  \\
      z_{-}  & \rightarrow & -e^{i\pi/6} \tau^z z_-
    \end{array}\right. ,\nonumber \\
  R_{2\pi/3} &:& \left\{ \begin{array}{lcr}  
      z_{+}  & \rightarrow & e^{-i \pi/3} u_r z_+  \\
      z_{-}  & \rightarrow & e^{i\pi/3} u_r z_-
    \end{array}\right. ,\nonumber \\
  I_{d_1} &:& \left\{ \begin{array}{lcr}  
      z_{+}  & \rightarrow & e^{i 5\pi/12} u_{\scriptscriptstyle 1}^{\vphantom*} z^*_+  \\
      z_{-}  & \rightarrow & e^{-i 5\pi/12} u_{\scriptscriptstyle 1}^{\vphantom*}z^*_-
    \end{array}\right. ,\nonumber \\
  I_{d_2} &:& \left\{ \begin{array}{lcr}  
      z_{+}  & \rightarrow & i u_{\scriptscriptstyle 2}^{\vphantom*} z^*_-  \\
      z_{-}  & \rightarrow & i u_{\scriptscriptstyle 2}^{\vphantom*}z^*_+
    \end{array}\right. , \nonumber \\
  C &:& \left\{ \begin{array}{lcr}  
      z_{+}  & \rightarrow & -\epsilon z_-^* \\
      z_{-}  & \rightarrow & \epsilon z^*_+
    \end{array}\right. .
\end{eqnarray}
Here
\begin{eqnarray}
  \label{eq:udef}
  u_r & = & e^{i \frac{\pi}{3} {\boldsymbol {\hat{n}}}_r\cdot
    {\boldsymbol\tau}} = 
  \frac{1}{2} \left( \begin{array}{cc} 1+i & 1+i \\ -1+i & 1-i
    \end{array}\right) ,\nonumber \\
  u_1 & = & e^{-i \frac{\pi}{2} {\boldsymbol {\hat{n}}}_1\cdot
    {\boldsymbol\tau}} = 
  \frac{1}{\sqrt{2}} \left( \begin{array}{cc} -i & -i \\ -i & i
    \end{array}\right) , \nonumber \\
  u_2 & = & e^{i \frac{\pi}{2} {\boldsymbol {\hat{n}}}_2\cdot
    {\boldsymbol\tau}} = 
  \frac{1}{\sqrt{2}} \left( \begin{array}{cc} -i & i \\ i & i
    \end{array}\right) ,
\end{eqnarray}
with
\begin{eqnarray}
  \label{eq:ndefs}
  {\boldsymbol{\hat{n}}}_r &  = & (1,1,1)/\sqrt{3}, \nonumber \\
  {\boldsymbol{\hat{n}}}_1 &  = & (1,0,1)/\sqrt{2}, \nonumber \\
  {\boldsymbol{\hat{n}}}_2 &  = & (1,0,-1)/\sqrt{2}.
\end{eqnarray}

Next we tabulate the transformation properties of the pseudospin
vector order parameters:
\begin{eqnarray}
  \label{eq:psv}
  T_1&:& \left\{ \begin{array}{lcr}  
      S_\alpha^x & \rightarrow & S_\alpha^x \\
      S_\alpha^y & \rightarrow & -S_\alpha^y \\
      S_\alpha^z & \rightarrow &  -S_\alpha^z
    \end{array}\right. ,\nonumber \\
  T_2&:& \left\{ \begin{array}{lcr}  
      S_\alpha^x & \rightarrow & -S_\alpha^x \\
      S_\alpha^y & \rightarrow & -S_\alpha^y \\
      S_\alpha^z & \rightarrow &  S_\alpha^z
    \end{array}\right. ,\nonumber \\
  R_{2\pi/3} &:& \left\{ \begin{array}{lcr}  
      S_\alpha^x & \rightarrow & S_\alpha^y \\
      S_\alpha^y & \rightarrow & S_\alpha^z \\
      S_\alpha^z & \rightarrow &  S_\alpha^x
    \end{array}\right. ,\nonumber \\
  I_{d_1} &:& \left\{ \begin{array}{lcr}  
      S_\alpha^x & \rightarrow & S_\alpha^z \\
      S_\alpha^y & \rightarrow & S_\alpha^y \\
      S_\alpha^z & \rightarrow &  S_\alpha^x
    \end{array}\right. ,\nonumber \\
  I_{d_2} &:& \left\{ \begin{array}{lcr}  
      S_\alpha^x & \rightarrow & -S_{-\alpha}^z \\
      S_\alpha^y & \rightarrow & S_{-\alpha}^y \\
      S_\alpha^z & \rightarrow &  -S_{-\alpha}^z
    \end{array}\right. , \nonumber \\
  C &:& {\bf S}_\alpha \rightarrow -{\bf S}_{-\alpha} .
\end{eqnarray}

\section{Hard spin tranformations}
\label{sec:hard-spin-tranf}

Here we give the transformation rules for the hard-spin theory with
parallel pseudospins.  Defining
\begin{equation}
  \label{eq:zpmapp}
  z_{\pm} = z e^{\pm i \theta/2},
\end{equation}
we obtain
\begin{eqnarray}
  \label{eq:hstfs}
   T_1&:& \left\{ \begin{array}{lcr}  
      z & \rightarrow & i \tau^x z \\
      \theta & \rightarrow &  \theta-4\pi/3
    \end{array}\right. ,\nonumber \\
   T_2&:& \left\{ \begin{array}{lcr}  
      z & \rightarrow & i \tau^z z \\
      \theta & \rightarrow &  \theta-4\pi/3
    \end{array}\right. ,\nonumber \\
   R_{2\pi/3} &:& \left\{ \begin{array}{lcr}  
      z & \rightarrow & u_r z \\
      \theta & \rightarrow &  \theta-2\pi/3
    \end{array}\right. ,\nonumber \\
   I_{d_1} &:& \left\{ \begin{array}{lcr}  
      z & \rightarrow & u_{\scriptscriptstyle 1}^{\vphantom*} z^* \\
      \theta & \rightarrow &  5\pi/6-\theta
    \end{array}\right. ,\nonumber \\
   I_{d_2} &:& \left\{ \begin{array}{lcr}  
      z & \rightarrow & u_{\scriptscriptstyle 2}^{\vphantom*} z^* \\
      \theta & \rightarrow &  \theta
    \end{array}\right. ,\nonumber \\
   C &:& \left\{ \begin{array}{lcr}  
      z & \rightarrow & \epsilon z^* \\
      \theta & \rightarrow &  \theta+\pi.
    \end{array}\right. \nonumber \\
\end{eqnarray}
The $SU(2)$ matrices $u$ are defined in Eqs.~(\ref{eq:udef}).  We have
used the freedom to redefine any of these operations by a global
$U(1)$ gauge transformation, which in this hard-spin limit corresponds
to a phase rotation of $z$.

\section{Residual symmetries of the SS2 supersolid}
\label{sec:resid-symm-ss2}

The residual symmetry operations of this SS2 state (which are not broken
by the pseudospin vector order) are generated by $T_1$, $R_{2\pi/3}\circ
I_{d_1}$, $C\circ I_{d_2} \circ I_{d_1}$, and $T_2\circ I_{d_2} \circ
I_{d_1}$.  Their action on the $z_\pm$ vortex fields can be obtained
from the definition $z_{\pm\sigma}=z_\pm \eta_\sigma$ and
Eqs.~(\ref{eq:ztfs}): 
\begin{eqnarray}
  \label{eq:zss3mott}
  T_1 & : &  z_\pm \rightarrow \pm e^{\mp i \pi/6} z_\pm, \nonumber \\
  R_{2\pi/3}\circ I_{d_1} & : &  \left\{ \begin{array}{ccc} z_+ &
      \rightarrow & z_+^*   \\ z_- &\rightarrow & i z_-^*
    \end{array}\right. \nonumber \\
  C\circ I_{d_2} \circ I_{d_1} & : & z_\pm \rightarrow e^{\mp i \pi/12}
  z_\pm^*, \nonumber \\
  T_2\circ I_{d_2} \circ I_{d_1} & : & z_\pm \rightarrow - e^{\pm i
    \pi/12} z_\mp.
\end{eqnarray}
These lead directly to Eq.~(\ref{eq:LSS2Mott}).

\section{Toy model wavefunctions with (111) pseudospin}
\label{sec:toy-model-wavefunctions}

\begin{figure}[htbp]
  \centering
  \includegraphics[width=4cm]{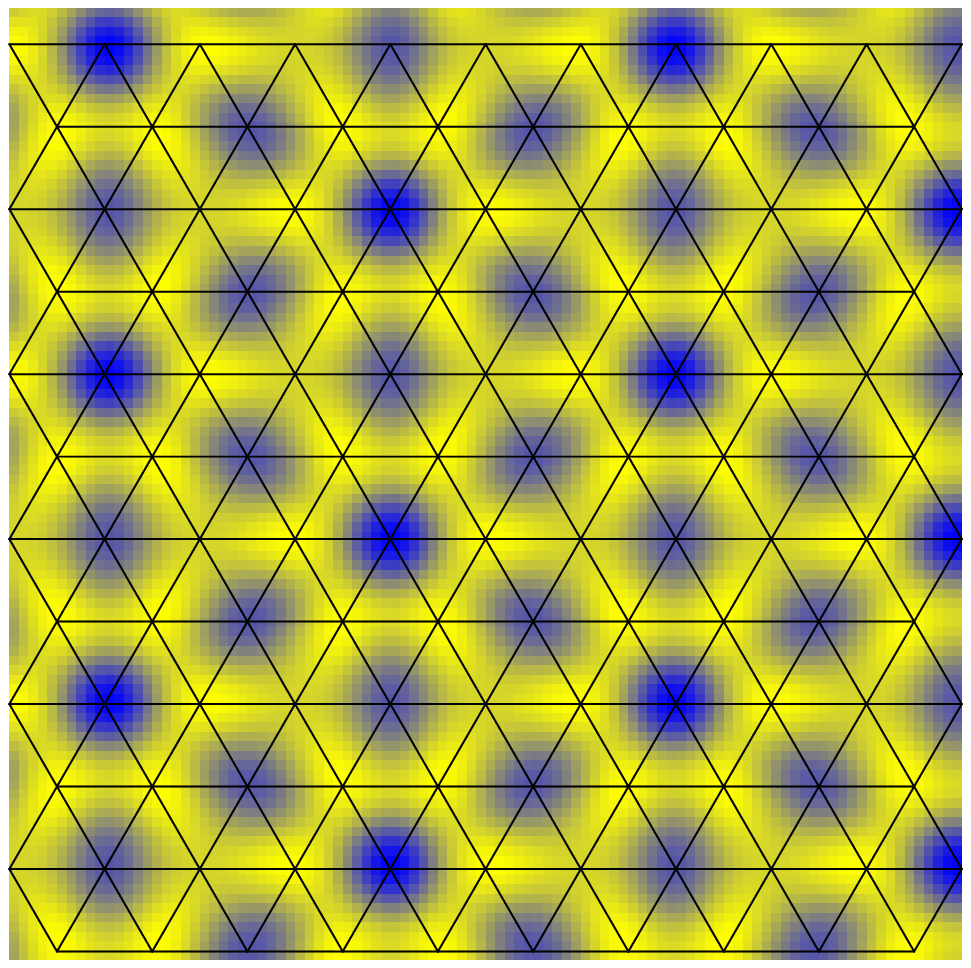}
\includegraphics[width=4cm]{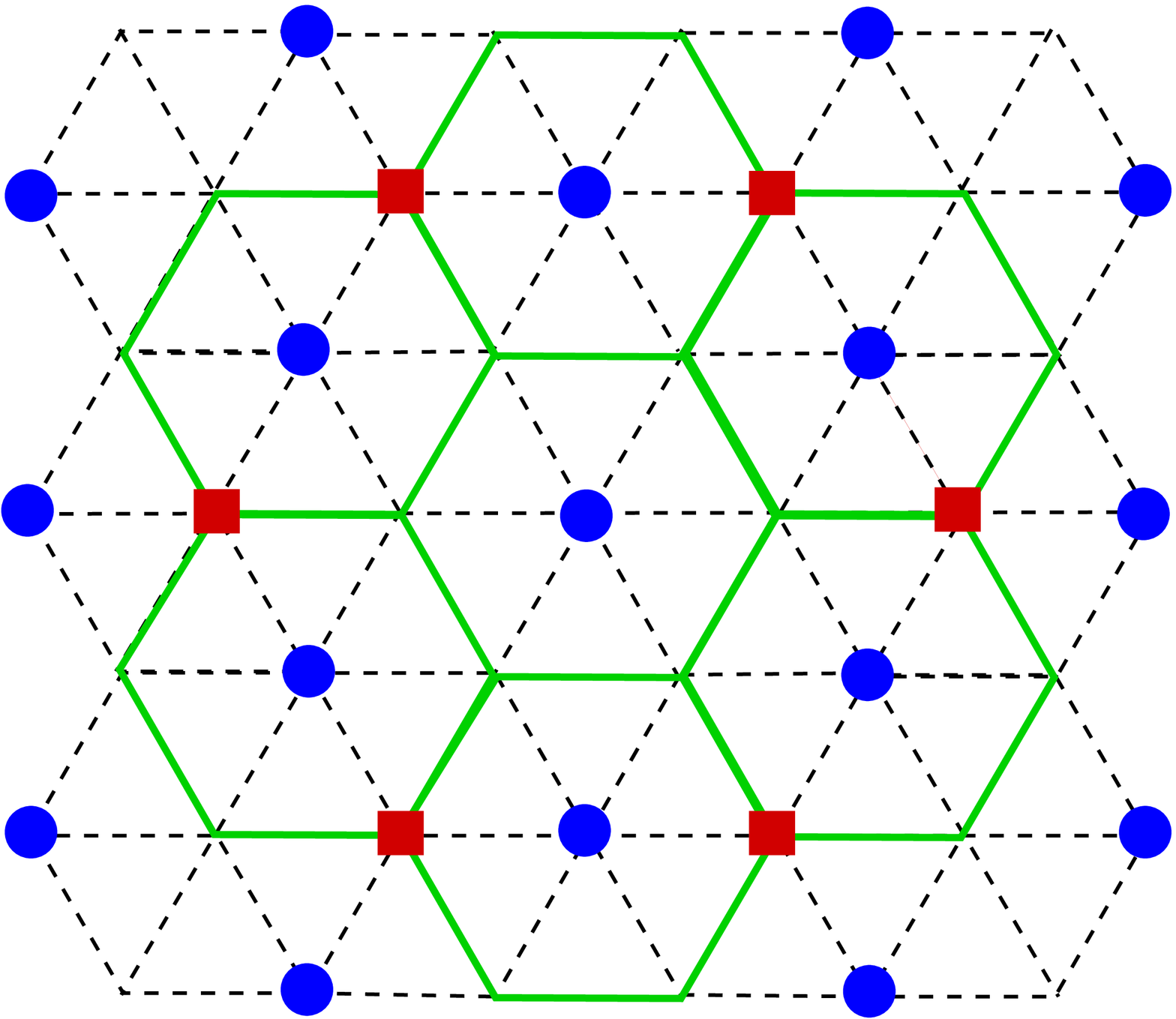}
  \caption{Density plot for the ferrimagnetic solid (left) and the corresponding cartoon (right) with (111) pseudospin ordering. }
\label{fig:ferrisolid111}
\end{figure}

Here we provide toy model wavefunctions for the Mott states with $w_2>0$
and parallel pseudospins, so that ${\bf S}_+={\bf S}_-$ lies along say
the $(111)$ axis.  First consider  $w_3>0$, in which we wish to
construct a state appropriate to a $1/4$-filled honeycomb lattice.
In this case, a wavefunction with the correct symmetry is shown in
Fig.\ref{fig:ferrisolid111}.  
In this figure the bosons denoted by squares are on the honeycomb sublattice 
of the triangular lattice, and become superfluid in the SS3 phase.

\begin{figure}[htbp]
  \centering
  \includegraphics[width=4cm]{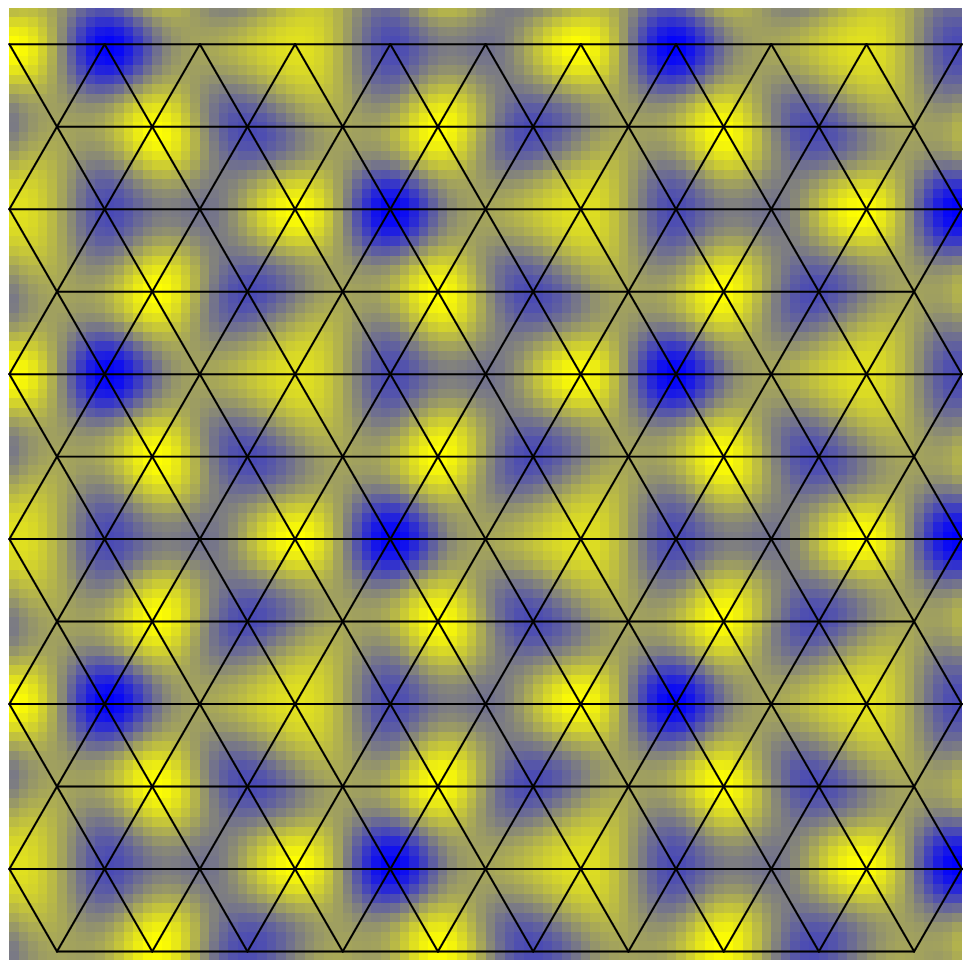}
  \includegraphics[width=4cm]{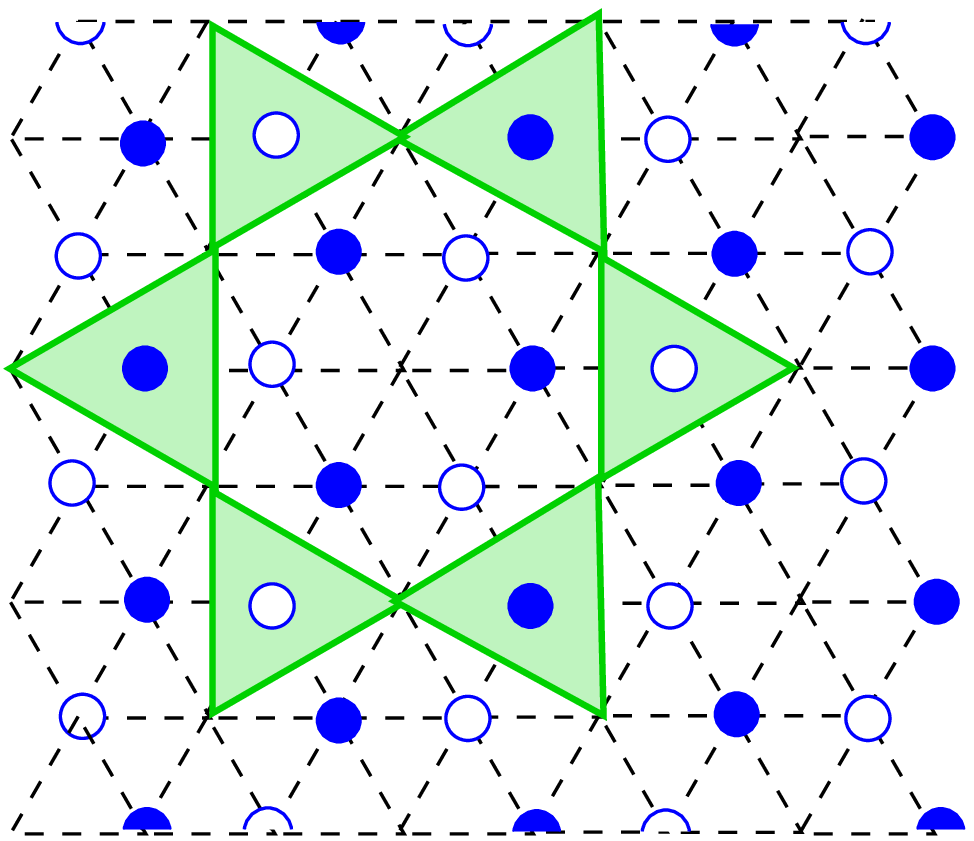}
  \caption{Density plot for the antiferromagnetic solid (left) and the corresponding cartoon (right) with (111) pseudospin ordering. }
  \label{fig:afsolid111}
\end{figure}

Now consider $w_3<0$, in which case we must consider the half-filled
triangular lattice embedded in the antiferromagnetic honeycomb Mott
insulator.  The density plot for this state is shown in 
Fig.\ref{fig:afsolid111}.
Focusing on the triangular lattice sites, we note that the amplitude
on these sites has the same $2\times 2$ periodicity as the $w_3>0$
state above, with the sites on this sublattice having one amplitude,
and the remaining sites another.  However, it is inconsistent with
half-filling to simply put one boson on the sublattice sites, or on
the other three sites.  Moreover, the state is rather constrained by
the existence of a number of centers of three-fold rotations and
reflections.  The simplest Mott wavefunction we constructed that
satisfies all symmetry properties is actually not quite a product
state of the form of Eq.~(\ref{eq:wf}).  It is, however, clearly an
insulating state.  To construct it, assume the $2\times 2$ sublattice
sites are empty.  Then the remaining sites form a {\sl
  kagome} lattice.  The kagome lattice is composed of corner-sharing
triangles, or two orientations (pointing ``left'' and ``right'' in the
figure).  We act with creation operators to place one boson on each
triangle in a uniform superposition.  
\begin{equation}
  \label{eq:psikag}
  \Psi = \prod_\triangle (b_{\triangle 1}^\dagger+b_{\triangle
    2}^\dagger+b_{\triangle 3}^\dagger ) |0\rangle
\end{equation}
Since the triangles overlap, there will actually be amplitude to find
more than one boson per triangle.  On average, the number of bosons
per unit cell of the kagome lattice is then $2$, since there is one up
and one down triangle per unit cell.  There are four triangular
lattice sites for each such unit cell (3 kagome and one central empty
site), so this is a state at $1/2$-filling, which manifestly has all
the symmetries of the Mott state in Fig.\ref{fig:afsolid111}.  
Moreover, it is an
insulating state as required, since one can readily show there are
only very short-range correlations, e.g. in the boson Green's
function.

\section{Honeycomb lattice at $1/4$-filling}
\label{sec:honeycomb-lattice-at}

The vortex PSG's on the honeycomb lattice were worked out in Appendix~B
of the first paper in Ref.~\onlinecite{Balents04}.  They are sufficient to
develop a dual vortex theory of the superfluid-Mott transition on this
lattice.  For the case of $f=1/4$, there are two vortex flavors,
$\varphi_0,\varphi_1$, which can be combined into a spinor
$\varphi_\sigma$.  We consider the restrictions upon the vortex
Lagrangian by translations and rotations.  Suppressing the indices and
using Pauli matrices ${\boldsymbol \tau}_{\sigma\sigma'}$ to span the
spinor space, we have:
\begin{eqnarray}
  \label{eq:honeypsg}
  T_1 &:& \varphi \rightarrow \tau^x \varphi, \nonumber \\
  T_2 &:& \varphi \rightarrow \tau^z \varphi, \nonumber \\
  T_d &:& \varphi \rightarrow \tau^y \varphi, \nonumber \\
  R^{\rm dual}_{2\pi/3} & : & \varphi \rightarrow e^{-i\pi/3} u_r
  \varphi, 
\end{eqnarray}
where $u_r$ is given in Eq.~(\ref{eq:udef}).  Requiring invariance under
these PSG operations, one may readily write down a continuum Lagrangian
in terms of $\varphi$:
\begin{eqnarray}
  \label{eq:hclag}
  {\cal L}_{\rm honey} &=&\sum_{\sigma = 0,1}
  \left[|(\partial_{\mu} - i A_{\mu}) \varphi_{\sigma}|^2 + 
    s |\varphi_{\sigma}|^2\right] + u (\varphi_\sigma^*
    \varphi_\sigma^{\vphantom*})^2  \nonumber \\ 
  &+&\frac{1}{2 e^2} 
(\epsilon_{\mu \nu \lambda}\partial_{\nu} A_{\lambda})^2 + {\cal L}_a,
\end{eqnarray}
where $SU(2)$ symmetry of $\varphi_\sigma$ is first broken by the term
\begin{equation}
  \label{eq:Lahon}
  {\cal L}_a = \lambda S^x S^y S^z,
\end{equation}
with ${\bf S} = \varphi_\sigma^* {\boldsymbol
  \tau}_{\sigma\sigma'}\varphi_\sigma$.  Eq.~(\ref{eq:hclag}) indeed has
the global $SU(2)$ and $U(1)$ gauge symmetry of the NCCP$^1$ model, and
${\cal L}_a$ recovers the leading anisotropy term in
Eq.~(\ref{eq:nccpanis}) for the ferrimagnetic case.

\end{document}